\documentclass{amsart}
\usepackage[utf8]{inputenc}
\usepackage{algorithmic} 
\usepackage{url}
\usepackage{amsmath,amssymb,amsthm}
\usepackage{graphicx}
\usepackage{fancyhdr}
\usepackage{xcolor}
\usepackage{hyperref}

\newcommand*{\hardyR}[1]{\mathsf{H}^{#1}({\mathbb R})}
\newcommand*{\blp}{{\mathsf B}}
\newcommand*{\cnj}[1]{\overline{#1}}

\begin{document}

\title[]{On the Practical Use of Blaschke Decomposition in Nonstationary Signal Analysis}
\author{Ronald R. Coifman}
\address{Department of Mathematics, Program in Applied Mathematics, Yale University, New Haven, CT 06510}
\email{coifman-ronald@yale.edu}

\author{Hau-Tieng Wu}
\address{Courant Institute of Mathematical Sciences, New York University, New York, NY, 10012}
\email{hauwu@cims.nyu.edu}

\maketitle

\begin{abstract}
The Blaschke decomposition-based algorithm, {\em Phase Dynamics Unwinding} (PDU), possesses several attractive theoretical properties, including fast convergence, effective decomposition, and multiscale analysis. However, its application to real-world signal decomposition tasks encounters notable challenges. In this paper, we propose two techniques, divide-and-conquer via tapering and cumulative summation (cumsum), to handle complex trends and amplitude modulations and the mode-mixing caused by winding. The resulting method, termed {\em windowed PDU}, enhances PDU's performance in practical decomposition tasks. We validate our approach through both simulated and real-world signals, demonstrating its effectiveness.   
\end{abstract}

\section{Introduction}

In the past decades, thanks to rapid innovation in technology, a proliferation of biosensors providing various biomedical signals. Compared with the {\em snapshot} health information, like that from the chest X-ray, questionnaire, or blood test, the information contained in time series is about dynamics \cite{benchetrit2000breathing,lombardi2000chaos,ahmad2009clinical}. Biomedical time series are inherently nonstationary and typically consist of multiple oscillatory components, each corresponding to an observation of a physiological system and characterized by time-varying amplitude, frequency, and nonsinusoidal waveforms. For example, the PPG signal \cite{kyriacou2021photoplethysmography} is composed of not only the volume information as the trend, but also the cardiac oscillatory component and the respiratory component \cite{shelley2007photoplethysmography}. A key mission in biomedical signal processing is {\em decomposing} it into each individual ingredient so that different components can be analyzed for different purposes. However, achieving this decomposition mission, or analyzing the dynamics, is in general challenging.

In this paper, we consider the {\em Phase Dynamics Unwinding} (PDU) algorithm, which is an iterative algorithm originally introduced in \cite{Nahon:2000Thesis} with Blaschke decomposition in each step, to handle biomedical signals. PDU is a nonlinear extension of Fourier analysis\footnote{Sometimes, PDU is called nonlinear Fourier transform.} that incorporates the roots of analytic functions.
By modeling a signal as a holomorphic function on the unit circle $\partial\mathbb{D}\subset \mathbb{C}$, iterative step of PDU begins by de-meaning the input signal. The roots located inside the unit disk are then used to decompose the signal into its outer and inner components. The inner function characterizes the local and oscillatory behavior within the current iteration, while the outer function serves as the input for the next iteration, and the process continues until convergence.
Early numerical studies by Nahon demonstrated PDU's potential for signal decomposition and analysis. Since then, it has been applied in diverse contexts, including Doppler-based blood velocity measurements \cite{Healy:2009a, Healy:2009b},  high-quality ECG compression \cite{tan2018novel}, heartbeat classification \cite{tan2019novel}, and phase analysis of biomedical signals \cite{huang2020reconsider}. PDU has also been used for root localization of analytic functions and its application to analyze gravitational wave data is given in \cite{Coifman_Steinerberger_Wu:2016}. Beyond applications, PDU's theoretical foundations have been actively explored over the past decade. We summarize these developments in the following.

Although the PDU algorithm enjoys a strong and elegant theoretical foundation rooted in complex analysis, its practical applicability to real-world data remains limited. In practice, signal processing encompasses a variety of tasks, including but not limited to signal decomposition, compression and feature extraction. This paper focuses on investigating two limitations of PDU in signal decomposition and proposes strategies to overcome them.
First, due to the nature of Blaschke factorization, PDU struggles to accurately recover time-varying amplitudes (amplitude modulation, AM), especially when the AM cannot be well approximated by a polynomial. {See the left panel of Figure \ref{windowBKD} for an example.} To address this, we introduce a divide-and-conquer approach inspired by time-frequency analysis: we segment the signal using a suitable windowing scheme, apply PDU locally to each segment, and then stitch the results to reconstruct the full decomposition. The philosophy is that, locally, the AM can be well approximated by a polynomial so that PDU works.
Second, when high-frequency components dominate in amplitude, the unwinding dynamics can become misleading, compromising the decomposition. {See the top left panel of Figure \ref{Nonsinusoidal:2} for an example.} This is rooted in the core mechanism of PDU, unwinding, which relies on the interaction between each oscillatory component and the origin. To mitigate this, we leverage the fundamental theorem of calculus, noting that integration acts as a Fourier multiplier of $1/\xi$, effectively suppressing high-frequency content and improving the robustness of the decomposition.

The paper is organized as follows. In Section~\ref{section:model}, we review the PDU algorithm, its numerical implementation, and the necessary mathematical background. In Section~\ref{sec: challenge and proposed solution}, we discuss the main challenges in signal decomposition and present the proposed techniques. Section~\ref{sec: simu and real data} demonstrates the performance of these methods on both simulated and real-world datasets. {Section~\ref{section: theoretical results} in the appendix summarizes existing theoretical results supporting the PDU framework for readers with interest in theory.} 
Throughout the paper, we systematically denote $\mathbb{D}:=\{z\in \mathbb{C}|\,\|z\|<1\}$ to be the unit disk in $\mathbb{C}$. For $p\geq 1$, denote $H^p:=H^p(\partial \mathbb{D})$ to be the Hardy space. Recall that $H^p(\partial \mathbb{D})$ can be identified with the subspace of $L^p(\partial \mathbb{D})$ whose Fourier coefficients of negative order vanish.

\section{Mathematical background and the PDU algorithm}\label{section:model}

Consider a holomorphic function $F$ over a disk of radius $1+\epsilon$ for any $\epsilon>0$ to model a given periodic analytic signal $f(t):[0,2\pi)\to \mathbb{C}$; that is, the Fourier coefficients $\hat{f}(n)=0$ for all $n<0$ and $f(t)=F(e^{it})$ for $t\in [0,2\pi)$.

\subsection{Review of Fourier series}
We start with a quick recast of the Fourier transform in the complex analysis language before reviewing the PDU.
For a holomorphic function $F:\mathbb{D} \rightarrow \mathbb{C}$, we can trivially rewrite it as
\[
F(z) = F(0) + (F(z) - F(0)).
\]
Since $0$ is a root of $F(z) - F(0)$, we have $F(z) - F(0) = z F_1(z)$ for another holomorphic $F_1$. By iterating the procedure, we generate one root at $0$ at a time and get
\begin{align}
F(z) &\,= F(0) + (F(z) - F(0)) 
= F(0) + z F_1(z)\nonumber \\
&\,= F(0) + z F_1(0) + z^2 F_2(0) + z^3 F_3(0) + \dots\,.\label{FourierSeriesExpansion}
\end{align}
If we set $z = e^{it}$, we have recovered the Fourier series of the function $F$ restricted to $\partial \mathbb{D}$ as
$$ 
f(t):=F( e^{it} ) =\sum_{k=0}^{\infty}{a_k e^{ikt}}.
$$
In short, Fourier series can be viewed as peeling off {\em one} root at $0$ of $F_l(z) - F_l(0)$ at the $l$th step iteratively.

\subsection{Review of vanilla PDU}

The key observation in \cite{Nahon:2000Thesis} is that we can ``unwind'' all zeroes of $F_l(z) - F_l(0)$ at once by factoring out  a ``normalized product'' of all zeroes.
This decomposition is a classical object in complex analysis known as the {\em Blaschke decomposition} \cite{Garnett:1981}, which states that 
any holomorphic $F:\mathbb{D} \rightarrow \mathbb{C}$ can be written as
$ F = B \cdot G$, where $B$ is the inner function of $F$ with the magnitude $1$ almost everywhere on $\partial \mathbb{D}$ and $G$ is the outer function of $F$ without roots inside the unit disk $\mathbb{D}$. Recall that $G(z):=\exp\{ \frac{1}{2\pi}\int_0^{2\pi}\frac{e^{it}+z}{e^{it}-z}\ln(|F(e^{it})|)dt\}$ and $B(z)=F(z)/G(z)$.
In our setup, $B$ is represented as a Blaschke product
satisfying $ B(z) = z^m \prod_{k}{\frac{z-\alpha_k}{1-\overline{\alpha_k}z}}$,
where $\alpha_k$ are roots of $F$ inside the unit disk.
An application of this fact repeatedly allows us to decompose the function $F$ as
\begin{align}
F(z) =&\, F(0) + (F(z) - F(0))\notag\\
=&\, F(0) + B_1(z) G_1(z)\notag\\
=&\, F(0) + B_1(z) (G_1(0) + (G_1(z)-G_1(0)))  \notag\\
=&\, F(0) + \sum_{k=1}^n G_k(0) \prod_{l=1}^k B_l(z) + \texttt{remainder}\label{PDUexpansion}\,,
\end{align}
where $F(z) - F(0)=B_1(z) G_1(z)$ {and $G_l(z)-G_l(0)=B_{l+1}(z)G_{l+1}(z)$ represent the successive Blaschke factorizations}. Note that we collect all roots inside the unit disk in $B_1$, so that $G_1$ is free of roots inside $\mathbb{D}$. Similarly for $G_l(z)-G_l(0)=B_{l+1}(z)G_{l+1}(z)$ for $l=1,2,\ldots$, while we note that $G_l(z)$ is free of roots inside $\mathbb{D}$ and we substract  $G_l(0)$ to generate at least one root inside $\mathbb{D}$. 
From \eqref{PDUexpansion}, define 
\[
F_0(z)=F(0)\quad\mbox{or}\quad f_0(t)=F(0),
\] 
where $t\in [0,2\pi)$, and for $\ell=1,2,\ldots$, define
\begin{align}
F_\ell(z):=G_\ell(0) \prod_{i=1}^\ell B_i(z)\quad\mbox{or}\quad f_\ell(t):=F_\ell(e^{it})=G_\ell(0) \prod_{i=1}^\ell B_i(e^{it})\,,\label{PDU positive frequency fl}
\end{align}
which we call the $\ell$-th factored {\em PDU-component}.

\subsection{Interpretation of PDU}

Clearly, PDU is a direct generalization of the Fourier series \eqref{FourierSeriesExpansion} in the sense that we take {\em all} roots of $G_l(z)-G_l(0)$ but not just $0$, 
and its has a very natural physical interpretation. First, $f_0(t):=F(0)$ is a constant function, which can be understood as the {\em constant trend}. $G_\ell(0)$ is also a constant function, which can be understood as the {\em constant amplitude} of the $\ell$-th PDU-component, and the phase of $\prod_{i=1}^\ell B_i(z)$ is monotonic so that its derivative describes the frequency of the $\ell$-th component. These facts come from a direct calculation. Suppose the Blaschke product $B$ contains $m$ roots at $0$ with other roots at $\alpha_k\in \mathbb{D}$ and can be expressed as
\begin{align}
B(e^{it}) = e^{imt} \prod_{k}{\frac{e^{it}-\alpha_k}{1-\overline{\alpha_k}e^{it}}}\,. 
\end{align}
Since $|B(e^{it})|=1$ for all $t\in [0,2\pi)$, if we set 
\begin{align}
B(e^{it}) = e^{i \phi(t)}, 
\end{align}
then by a direct calculation we have
\begin{align}
\phi'(t) = m + \sum_{k}\frac{1-|\alpha_k|^2}{|e^{i t} - \alpha_k|^2} > 0.\label{PDU positive frequency IF}
\end{align}
In particular, the phase $\phi(t)$ is always monotonically increasing, which means that $\phi'(t)$ is strictly positive and can be understood as the {\em instantaneous frequency} (IF). The capability to capture IF is one of the key strengths of PDU.

\begin{figure}[hbt!]
\includegraphics[trim=0 0 0 0, clip,width=0.495\textwidth]{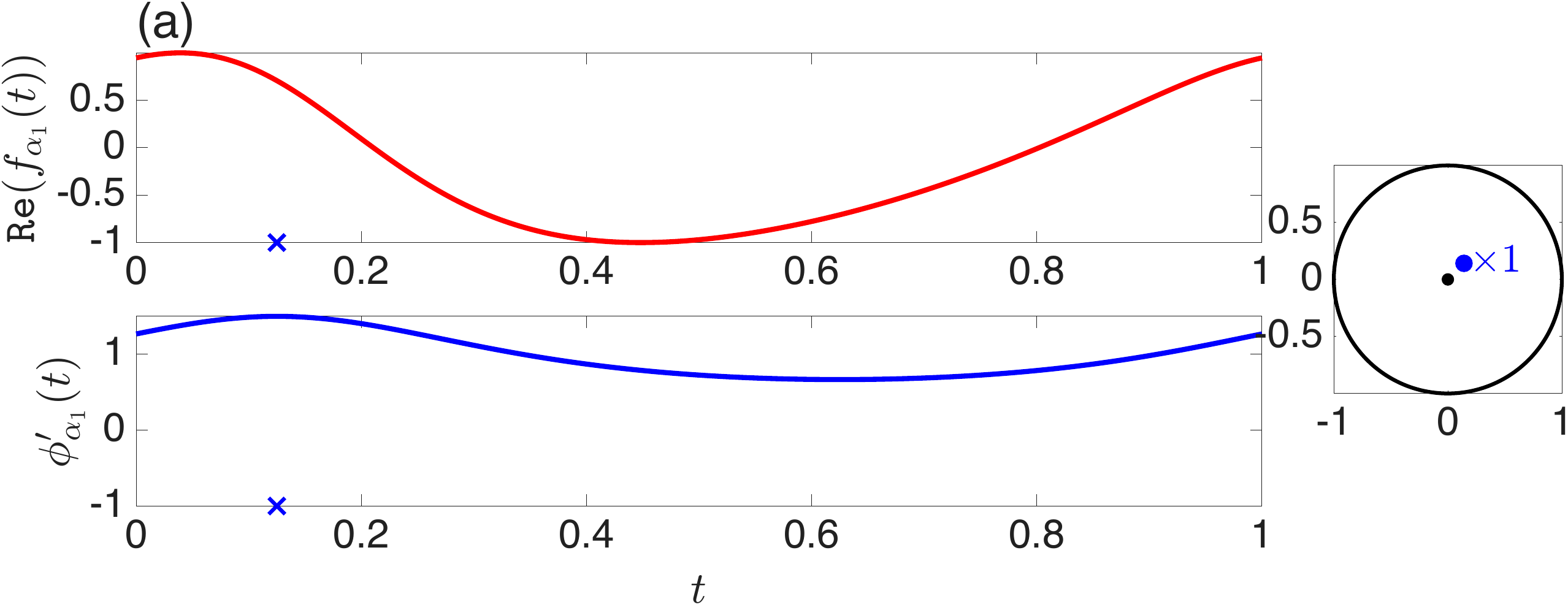}
\includegraphics[trim=0 0 0 0, clip,width=0.495\textwidth]{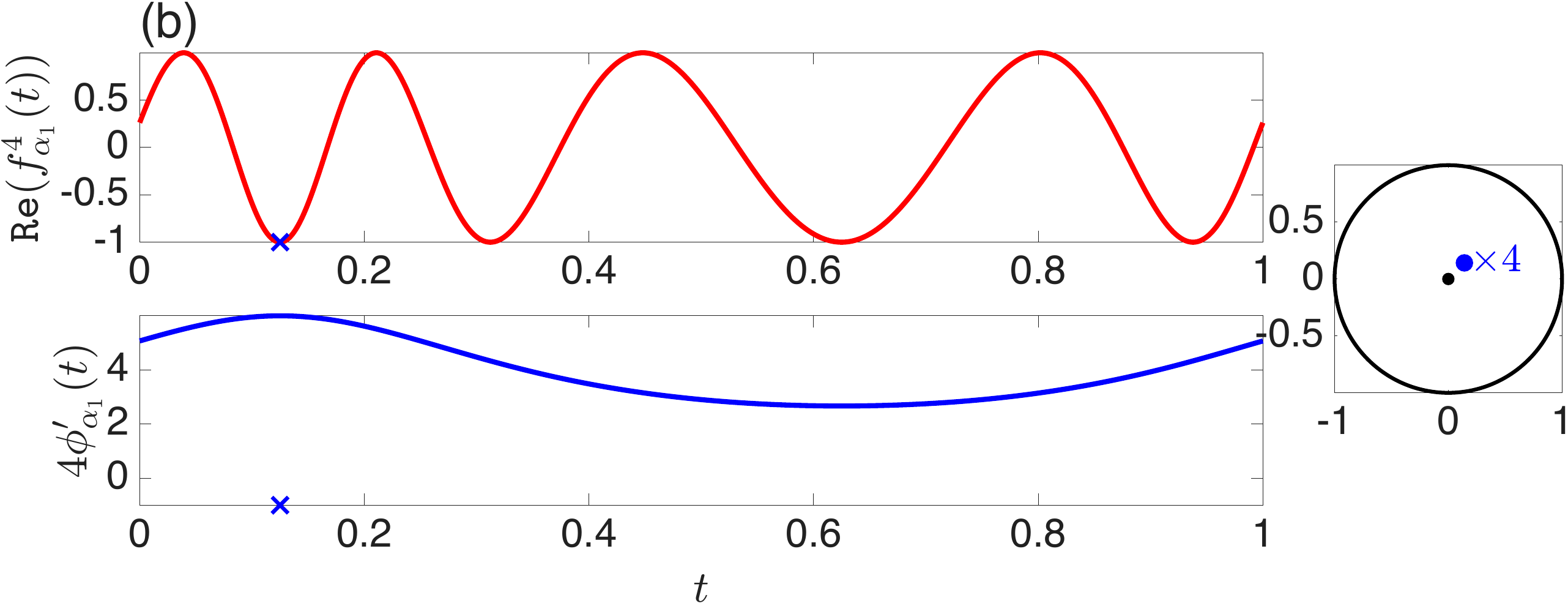}
\includegraphics[trim=0 0 0 0, clip,width=0.495\textwidth]{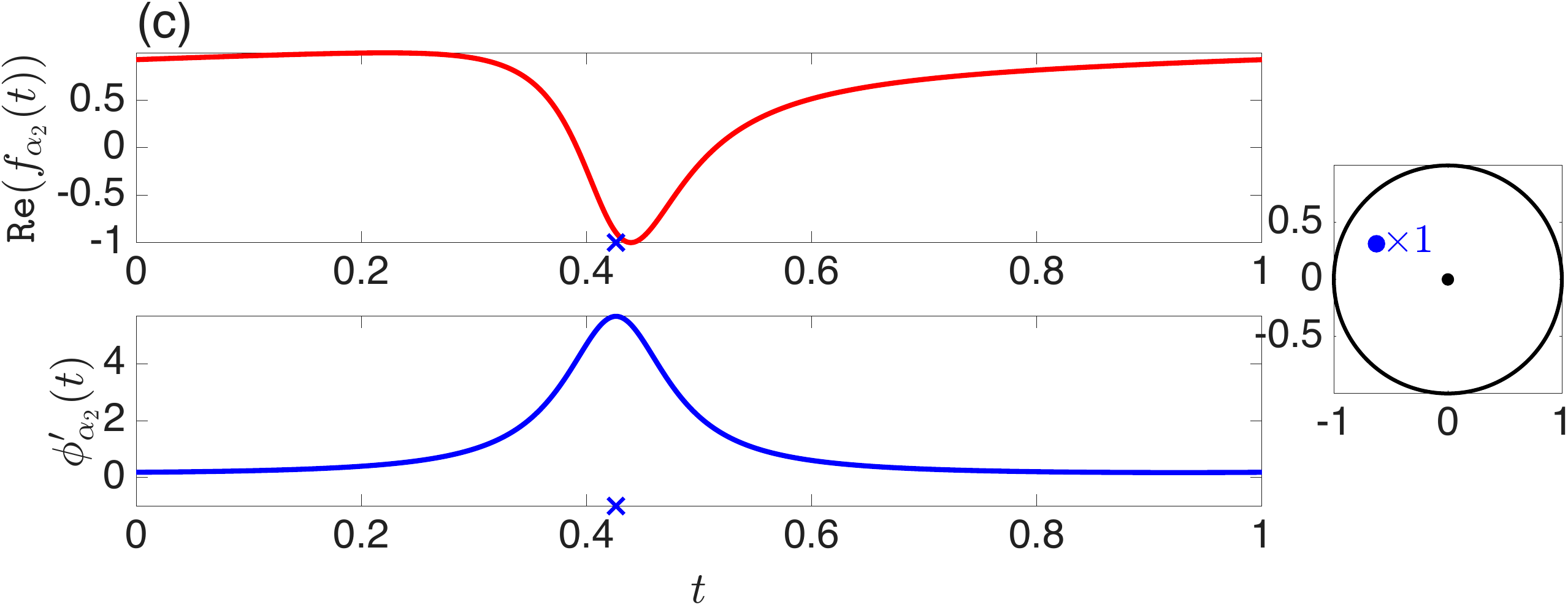}
\includegraphics[trim=0 0 0 0, clip,width=0.495\textwidth]{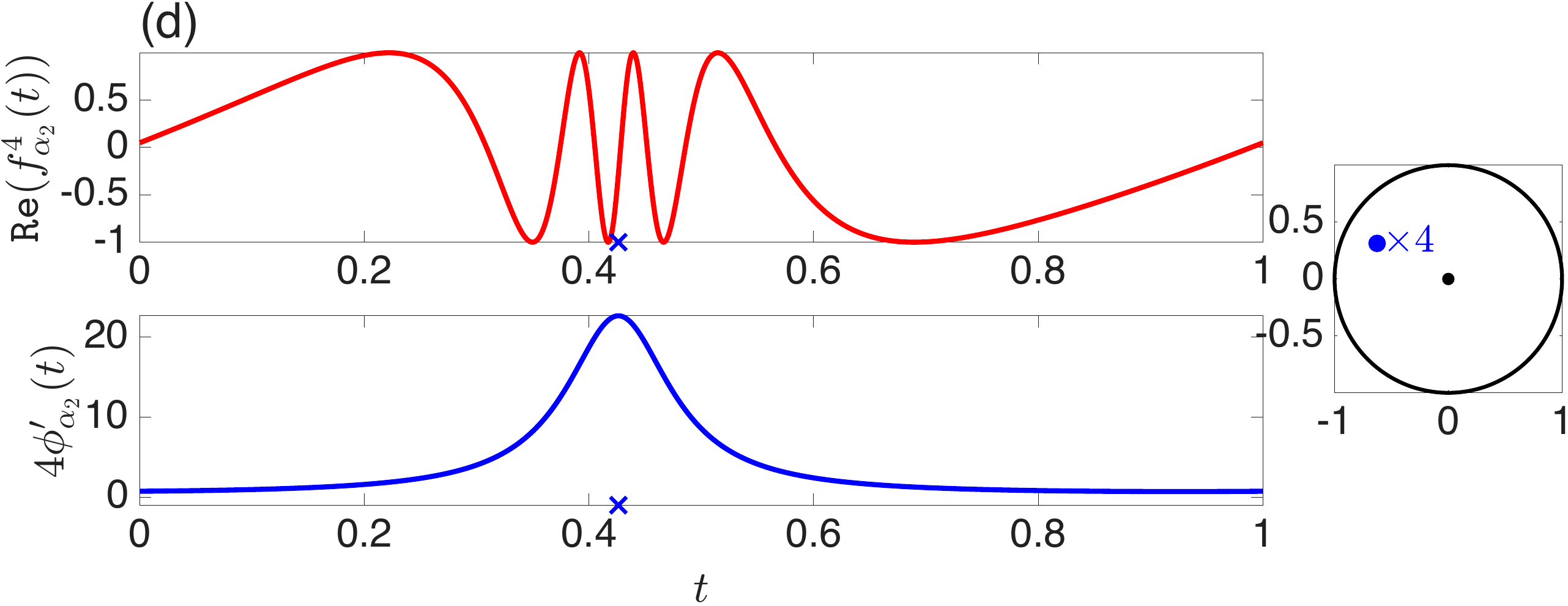}
\includegraphics[trim=0 0 0 0, clip,width=0.495\textwidth]{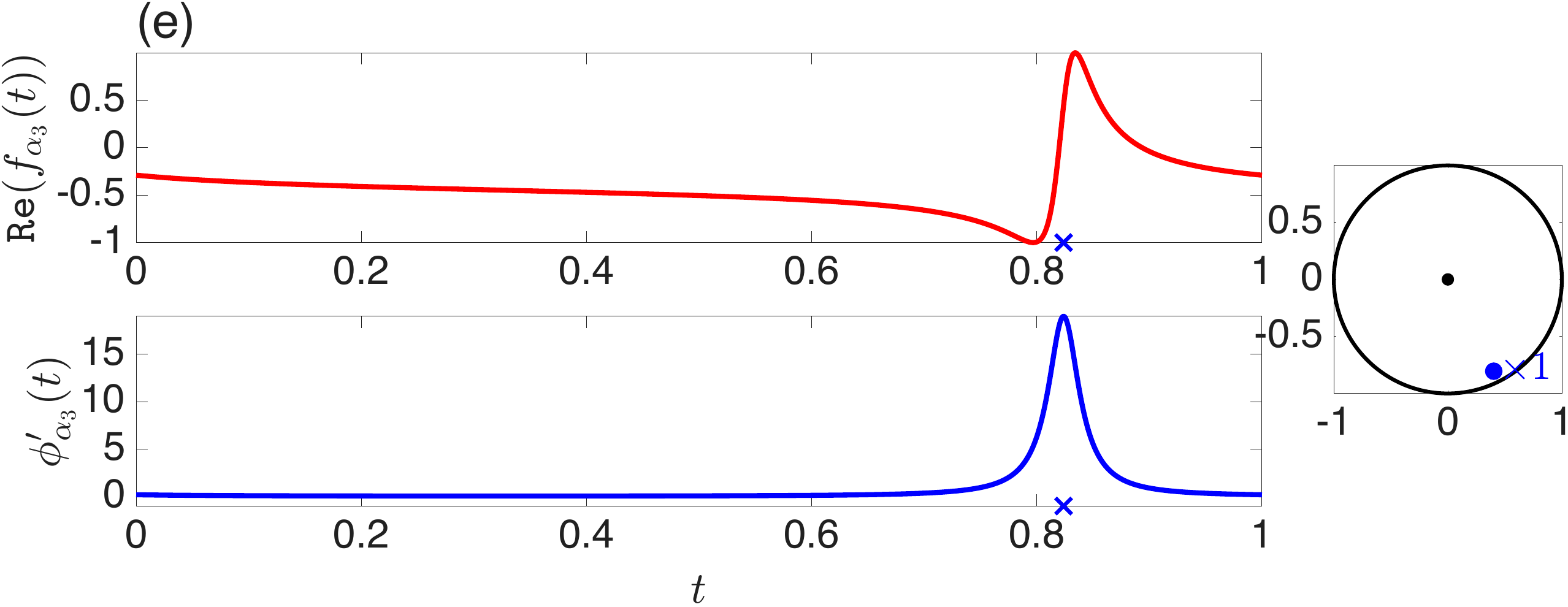}
\includegraphics[trim=0 0 0 0, clip,width=0.495\textwidth]{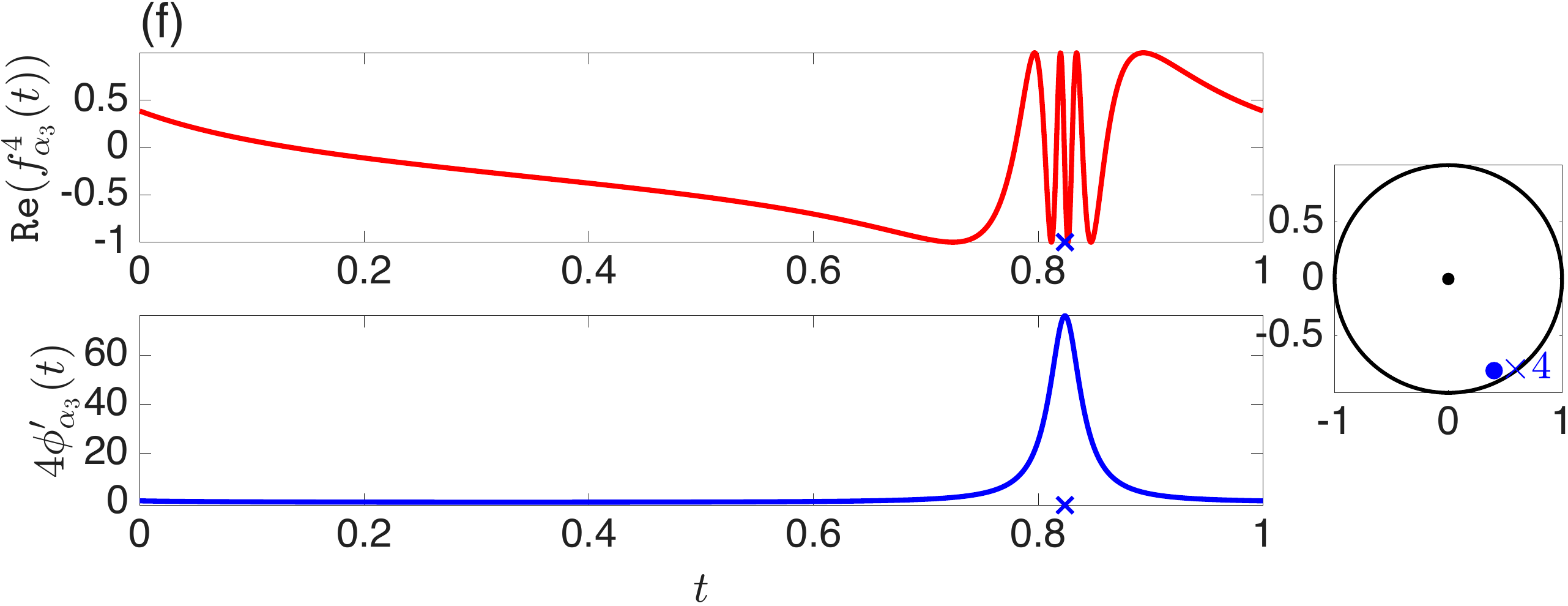}
\includegraphics[trim=0 0 0 0, clip,width=0.495\textwidth]{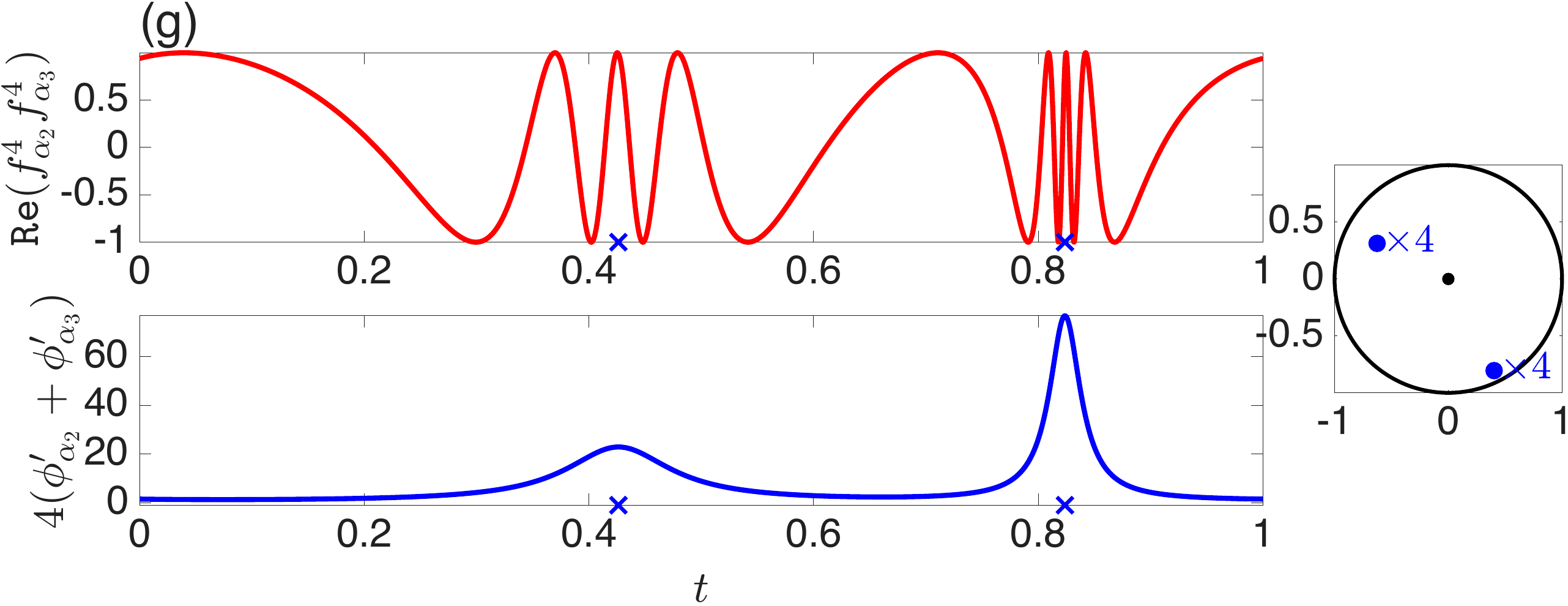}
\includegraphics[trim=0 0 0 0, clip,width=0.495\textwidth]{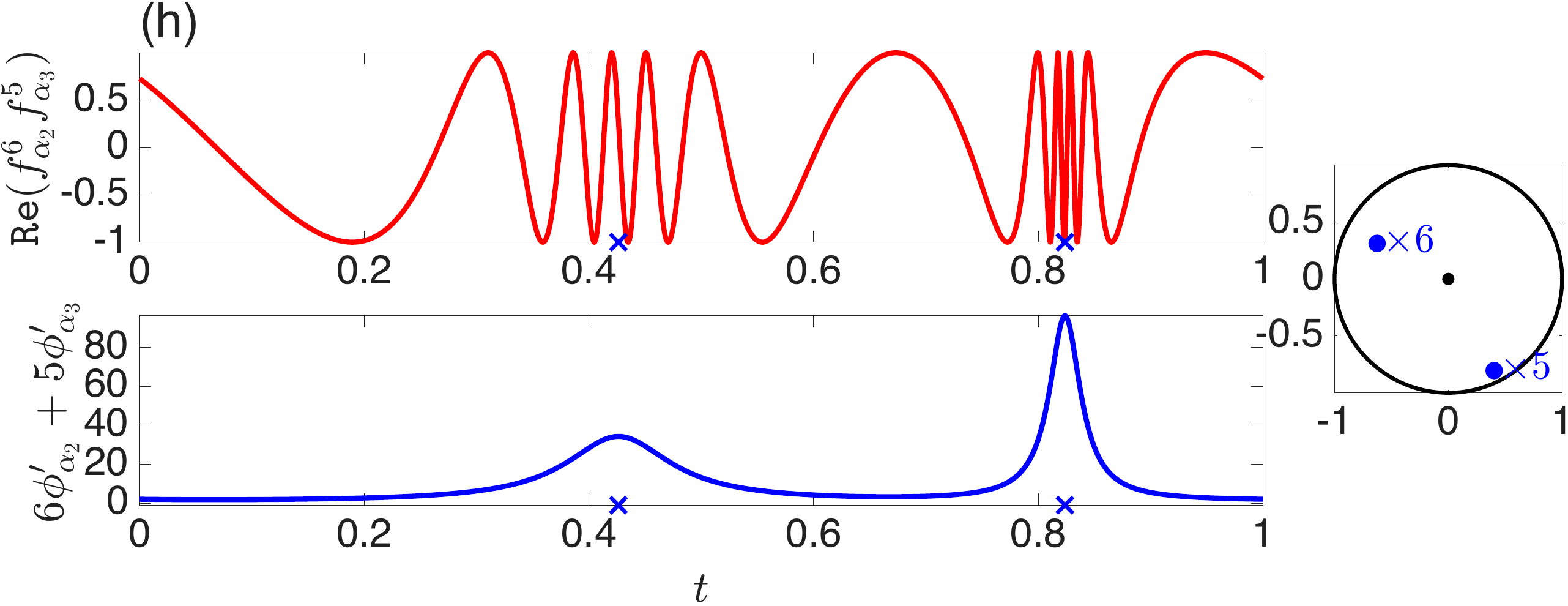}
\caption{Examples of the $f_\alpha(t):=\frac{e^{i2\pi t}-\alpha}{1-\overline{\alpha}e^{i2\pi t}}=e^{i\phi_\alpha(t)}$, where $t\in [0, 1)$, $\alpha=r_\alpha e^{i2\pi\theta_\alpha}\in \mathbb{D}$, $r_\alpha\in [0,1)$ and $\theta_\alpha\in [0,1)$. In each subplot, the red curve represents the real part of the signal, while the blue curve depicts the corresponding instantaneous frequency. The blue cross marks $\theta_\alpha\in [0,1)$, the phase of the associated root. In the right subpanel, the blue dot denotes the root located inside the unit disk, and the multiplicity of this root is indicated next to the blue cross.
(a) $f_{\alpha_1}$, where $\alpha_1=\frac{1+i}{5\sqrt{2}}$, (b) $f_{\alpha_1}^4$, 
(c) $f_{\alpha_2}$, where $\alpha_2=\frac{-14+7i}{10\sqrt{5}}$, (d) $f_{\alpha_2}^4$, 
(e) $f_{\alpha_3}$, where $\alpha_3=\frac{9-18i}{10\sqrt{5}}$, (f) $f_{\alpha_3}^4$, 
(g) $f_{\alpha_2}^4f_{\alpha_3}^4$, and (h) $f_{\alpha_2}^6f_{\alpha_3}^5$.  \label{fig Blaschke product example}}
\end{figure}

To understand how IF is encoded, consider a root $\alpha\in \mathbb{D}$. The term $\frac{1-|\alpha|^2}{|e^{i t} - \alpha|^2}$ is equal to $1$ when $\alpha=0$, indicating a constant frequency. When $\alpha=|\alpha|e^{i\theta}\neq 0$, $\frac{1-|\alpha|^2}{|e^{i t} - \alpha|^2}$ is large near $t=\theta$, suggesting a higher oscillation rate at time $\theta$. In other words, the root at 0 captures constant-frequency behavior, while roots closer to the boundary of the unit disk encode faster local oscillations. 
See Figure \ref{fig Blaschke product example} for an illustration. Clearly, the closer a root is to the unit circle, the faster the corresponding oscillation; and the more roots there are, the more rapid the oscillation. Therefore, $f_\alpha(t):=\frac{e^{i2\pi t}-\alpha}{1-\overline{\alpha}e^{i2\pi t}}$, where $t\in [0, 1)$, $\alpha=r_\alpha e^{i2\pi\theta_\alpha}\in \mathbb{D}$, $r_\alpha\in [0,1)$ and $\theta_\alpha\in [0,1)$, can be viewed as a ``template'' of a single oscillation localized around time $\theta_\alpha$. Note that the magnitude $r_\alpha$ also determines the ``temporal localization'' or ``spread'' of this oscillation, where smaller values of $r_\alpha$ correspond to broader support, while values closer to $1$ result in more sharply localized oscillations. Indeed, a direct calculation shows that the peak IF is $\frac{1+r_\alpha}{1-r_\alpha}$, attained at time $\theta_\alpha$. This indicates that when $r_\alpha$ is close to $1$, the oscillation becomes highly ``concentrated'' within an interval of length approximately $1-r_\alpha$. Raising this function to the $n$-th power, for any $n\in \mathbb{N}$, produces $n$ repetitive oscillations within the same localized subset. In this way, the widely studied quantity, IF, is naturally encoded in the roots of the Blaschke product. Take $f_{\alpha_3}$ as an example and assume the time unit is seconds. The ``single oscillation'' occurring around $t=0.8$ s lasts approximately $0.06$ s,  suggesting an IF of about 16 Hz; that is, roughly 16 oscillations per second, but only over a short interval. Intuitively, the IF of $f_{\alpha_3}^4$ is much higher near $t=0.8$ s, as each ``oscillation'' completes in roughly $1/4$ period of time. Note that over one quarter of the original period. In contrast, over the intervals $[0,0.7]$ s and $[0.9,1]$ s, the signal remains essentially constant with no visible oscillations, which the IF captures as values close to, but slightly above, zero. This quantification aligns well with our physical intuition.

{
We shall also mention that from Fourier analysis perspective, $f_\alpha$ is of broad spectrum when $\alpha\neq 0$. Indeed, by a direct binomial expansion, we have $\widehat{f_{\alpha}}(k)=-\alpha$ when $k=0$, $\widehat{f_{\alpha}}(k)=(1-|\alpha|^2)\bar{\alpha}^{k-1}$ when $k>0$, and $\widehat{f_{\alpha}}(k)=0$ when $k<0$. Moreover, for $f_\alpha^n$, where $n\in \mathbb{N}$, it has been known that as $n$ grows, the low frequency Fourier coefficients of $f_\alpha$ exponentially decay to $0$ with the rate depending on $\alpha$ \cite{borichev2024fourier}. This result again explains the nonlinear relationship between the notion of frequency from the Fourier perspective and instantaneous frequency.
}

Signal processing tasks can be performed once the PDU-components are obtained. For signal decomposition, we expect each $f_\ell$ to capture meaningful individual components. For compression, the goal is to use as few PDU-components as possible to reconstruct the original signal $F$. For feature extraction in machine learning applications, we can use either the roots or the components $f_\ell$ as informative features.

\subsection{Review of PDU with low pass filter to capture amplitude modulation}\label{section: review PDU LFP}

The capability of capturing IF is the main strength of PDU. However, the trend and amplitude modulation of all PDU-components are constant. Indeed, since inner functions are bounded analytic functions with modulus $1$ on $\partial \mathbb{D}$, a component like $F_3(z)=G_3(0) B_1(z) B_2(z) B_3(z)$ will always have unit amplitude. 
From this perspective, we immediately encounter a limitation of the approach in the context of signal decomposition, which restricts vanilla PDU's ability to obtain decomposed components with realistic trend and amplitude modulation (AM), limiting its practical applications.

{ Qualitatively, AM describes the strength of an oscillatory signal, while the trend refers to a slowly varying component. For example, for each PDU-component, its magnitude on $\partial \mathbb{D}$ is $1$, and hence we say it has AM equal to $1$. However, in practice an oscillatory component can exhibit time-varying strength, which a single PDU-component does not capture.
To discuss meaningful decomposition via Blaschke (vanilla PDU), we adopt one defintion of AM under the adaptive harmonic model (AHM) \cite{DaLuWu2011}.
Assume the input signal is analytic and satisfies 
\begin{equation}\label{model: AHM equation}
f(t)=\sum_{l=1}^K A_l(t)e^{2\pi i\phi_l(t)}+T(t)\,, 
\end{equation}
where $t\in [0,2\pi)$, or $\partial \mathbb{D}$, each $A_l(t)$ is positive smooth function with $|A_l'(t)|\leq \epsilon A_l(t)$ for some $\epsilon\geq 0$, $\phi_l(t)$ is smooth monotonically increasing function with $|\phi''_l(t)|\leq \epsilon \phi'_l(t)$ and $\phi'_l(t)-\phi'_{l-1}(t)>\Delta>0$ for $l=2,\ldots,K$ and $\phi'_1(t)>\Delta$ for some $\Delta>0$, and $T(t)$ is a slowly varying smooth function whose Fourier transform is dominated in a small neighborhood of $0$. 
We call $A_l(t)$ and $\phi'_l(t)$ the AM and IF of the $l$-th intrinsic mode type (IMT) function $g_l(t):=A_l(t)e^{2\pi i\phi_l(t)}$, and $T(t)$ the {\em trend}. Here, $\epsilon$ quantifies how fast the AM and IF can vary. The goal is to decompose $f(t)$ into $g_l(t)$, $l=1,\ldots,K$ and $T(t)$ by iteratively applying Blaschke decomposition, or vanilla PDU.
See Appendix for more discussion about AHM in the general case. Clearly, vanilla PDU cannot accurately decompose such signal.}

To resolve this limitation, a solution was proposed in \cite{Nahon:2000Thesis}. The key observation is that $F(0)$ ($G_\ell(0)$ respectively) is the mean, or the 0th order Fourier series coefficient of $F$  ($G_\ell$ respectively). In \cite{Nahon:2000Thesis}, the author proposed to remove a low frequency component in each step instead to capture the potentially non-constant trend and AM; that is, consider
\begin{align}
F(z) =&\, \mathcal{F}_LF(z) + (F(z) -\mathcal{F}_L F(z))\notag\\
=&\, \mathcal{F}_LF(z) + B_1(z) G_1(z)\notag\\
=&\, \mathcal{F}_LF(z) + B_1(z) (\mathcal{F}_LG_1(z) + (G_1(z)-\mathcal{F}_LG_1(z)))  \notag\\
=&\, \mathcal{F}_LF(z) + \sum_{k=1}^n \mathcal{F}_LG_k(z) \prod_{l=1}^k B_l(z) + \texttt{remainder}\label{PDUexpansion2}\,,
\end{align}
where $L\in \mathbb{N}\cup\{0\}$ is the order of low pass filter; that is, $\mathcal{F}_LF(z)=\sum_{l=0}^L \hat{f}(l)z^l$. In other words, in each step we generate different roots inside $\mathbb{D}$ compared with the vanilla PDU, and $0$ is still included. Clearly, when $L=0$, $\mathcal{F}_0F(z)=F(0)$ and we recover the vanilla PDU. 
Similarly, define $F_0(z)=:\mathcal{F}_LF(z)$ and $f_0(t):=\mathcal{F}_LF(e^{it})$ for $t\in [0,2\pi)$, and for $\ell=1,2,\ldots$, define the $\ell$-th factored {\em PDU-component} by
\begin{align}
F_k(z):=\mathcal{F}_LG_k(z) \prod_{l=1}^k B_l(z)\quad\mbox{and}\quad f_k(t):=\mathcal{F}_LG_k(e^{it}) \prod_{l=1}^k B_l(e^{it})\,.\label{PDU positive frequency fl2}
\end{align}
We refer to this algorithm as PDU to distinguish it from the vanilla PDU. In this setup,  $\mathcal{F}_LF(z)$ is called the trend, while $\mathcal{F}_LG_k(z)$ represents the AM of $F_k(z)$. 

{Consider an example $f(z)=\sum_{l=1}^Lc_lz^l+f_\alpha^n(z)$, where $c_l\in \mathbb{C}$, $\sum_{l=1}^Lc_lz^l$ models the trend, and $f_\alpha(z)=\frac{z-\alpha}{1-\bar{\alpha}z}$, $\alpha\in \mathbb{D}$ and $n> L$, models the oscillatory component. With the above discussion, when $n$ is sufficiently large, the spectrum of $f_\alpha^n$ is away from $0$, so that the low pass filter of order $L$ can decompose the ``trend'' $cz$ and the oscillatory component $f_\alpha^n$ with high precision.}

While this idea successfully capture the non-constant trend and AM, polynomial is too limited to capture complex trend and AM behavior {like that in \eqref{model: AHM equation}}. The main focus of this paper is introducing an idea to handle this challenge, which will be detailed below.

\subsection{Numerical Implementation of PDU}\label{Section: numerical implementation of PDU}

{In practice, a given signal is usually real-valued, non-periodic, and discretized. Before applying PDU, we apply the following pre-processing if necessary. Extend the signal by flipping and stitching; that is, for a given signal $\mathbf f_0\in \mathbb{R}^N$ sampled from a continous signal $f$ defined on $[0,L]$, extend it to $\mathbf f\in \mathbb{R}^{2N}$ by setting $\mathbf f(l)=\mathbf f_0(l)$ for $l=1,\ldots,N$, and $\mathbf f(L+l)=\mathbf f_0(L-l+1)$ for $l=1,\ldots,N$. As a result, the extended $\mathbf f$ is a discretization of a continuous function on $[0,2L]$ that comes from flipping and stitching $f$, or via a dilation it is a discretization of a continuous function on $[0,2\pi)$, or $\partial\mathbb{D}$. Then, apply the discrete Hilbert transform to convert $\mathbf f$ to a holomorphic one.}

Surprisingly, while theoretically we need to know all roots of $G_l(z)-G_l(0)$ for the $l$th step, which is numerically challenging, in practice this step is not needed, thanks to the theoretical result reported in \cite{Weiss_Weiss:1962}. 
The result states that since $|F|=|G|$ on $\partial\mathbb{D}$ and $\ln(G)$ is analytic in the disk, by denoting 
\[
P_+(\ln|F|):=\sum_{k=0}^{\infty}\mathcal{F}(\ln|F|)(k)  e^{ikx}, 
\]
we have that
$$
G=\exp(P_+(\ln|F|))\quad\mbox{and}\quad B=F/G.$$
This theoretical result leads to an efficient numerical implementation. 

{With the preprocessed signal $\mathbf{f}$,} apply the discrete Hilbert transform to $\ln(|\mathbf{f}|+\epsilon)$, where $\epsilon>0$ is a small constant to control the blowup of $\ln$ when any entry of $|\mathbf{f}|$ is close to zero. Denote the result as $\mathbf{h}\in \mathbb{R}^{2N}$. Due to the continuity, the discretization of $G(e^{i\theta})$, denoted as $\mathbf{G}\in \mathbb{C}^{2N}$ is nothing but applying exponential to $\mathbf{h}$ entrywisely, and the discretization of $B(e^{i\theta})$, denoted as $\mathbf{B}\in \mathbb{C}^{2N}$, is the entrywise division of $\mathbf{f}$ by $\mathbf{G}$. 
{For the low pass filter in PDU, we may consider $L=5$. Repeat the above numerical Blaschke decomposition iteratively and obtain a sequence of components, denoted as $\mathbf{f}_1,\mathbf{f}_2,\ldots\in\mathbb{C}^{2N}$. The real parts of $\mathbf{f}_1,\mathbf{f}_2,\ldots$, restricted on the first half of the samples when the extension step is used, constitute the resulting decomposition.}

When the original signal is undersampled or has rapid variations, the discrete Hilbert transform can become inaccurate due to aliasing and poor frequency resolution. Therefore, in practice it is usually suggested to upsample the signal to increases time resolution and allow for better numerical approximation of the Hilbert transform and hence the phase and amplitude recovery of the analytic signal. Upsampling also reduces phase aliasing and prevents artificial phase jumps when unwrapping the phase, leading to more stable Blaschke decomposition.
Since Fast Fourier transform is the main algorithm that PDU is based on, it is numerically efficient, even after upsampling.

A variety of numerical experiments carried out in \cite{Nahon:2000Thesis}, and various applications afterward, suggest several nice practical properties of PDU.
Empirically, the convergence rate {of the PDU decomposition appears to be exponential, since only a few terms suffice to recover the input signal with a small $L^2$ reconstruction error.} This property has been utilized to design a high quality electrocardiogram (ECG) compression algorithm \cite{tan2018novel}.
Moreover, the PDU decomposition appears to be stable under small, structured perturbation, which is particularly useful when the signal contains segments of constant values, or so-called ``silent period''. Specifically, note that when there is no fluctuation, the PDU algorithm can becomes unstable since the holomorphic structure is disrupted by the presence of roots on open subsets of $\partial\mathbb{D}$. Such situations are commonly encountered in practice. Beyond a straightforward approach that detects and removes these segments before analyzing the remaining parts separately, one may instead add a small-amplitude sinusoidal perturbation and then apply the PDU algorithm, as suggested in \cite{Saito_Letelier:2009}. These properties are partially backed up by existing theoretical results, which {are summarized in Section~\ref{section: theoretical results}.}

\section{Challenges and proposed techniques}\label{sec: challenge and proposed solution}

\begin{figure}[hbt!]
\includegraphics[trim=0 0 0 0, clip,width=0.495\textwidth]{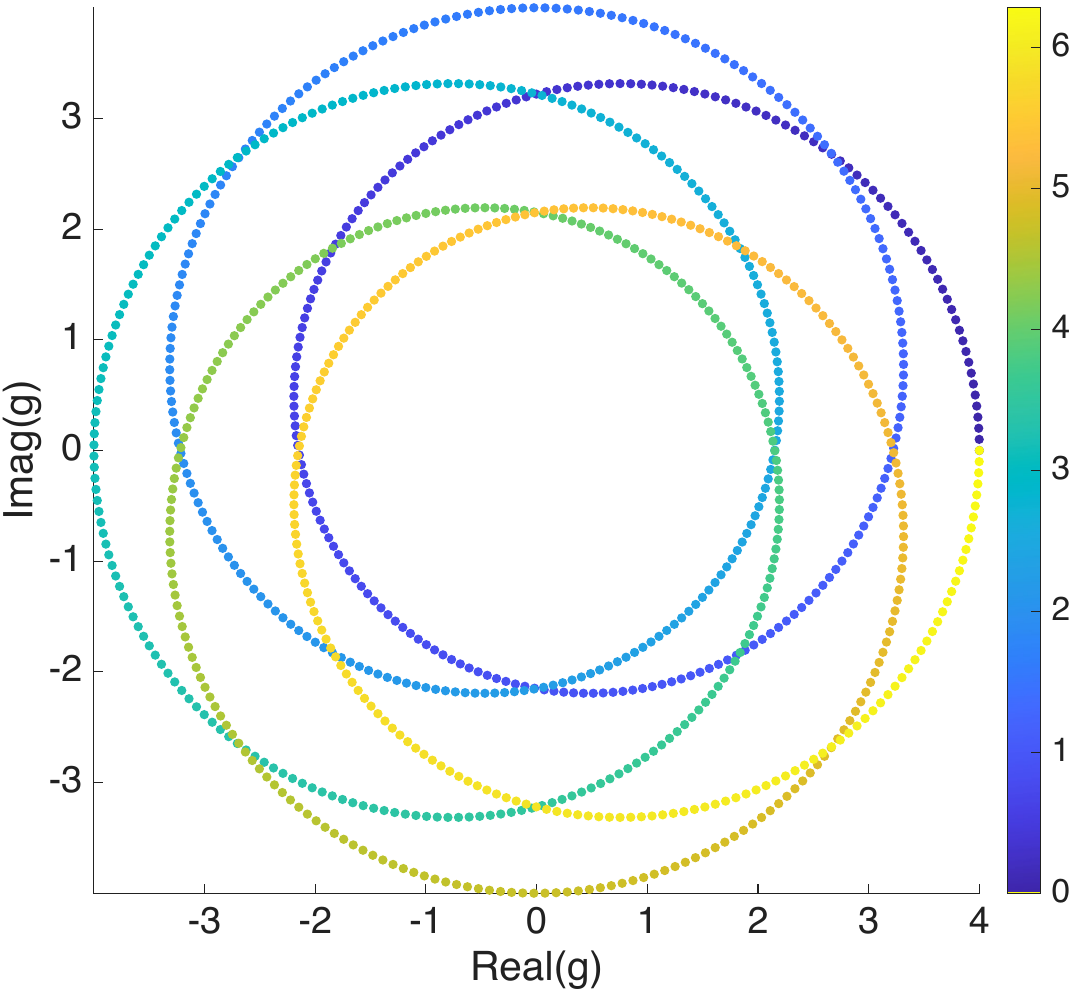}
\includegraphics[trim=0 0 0 0, clip,width=0.495\textwidth]{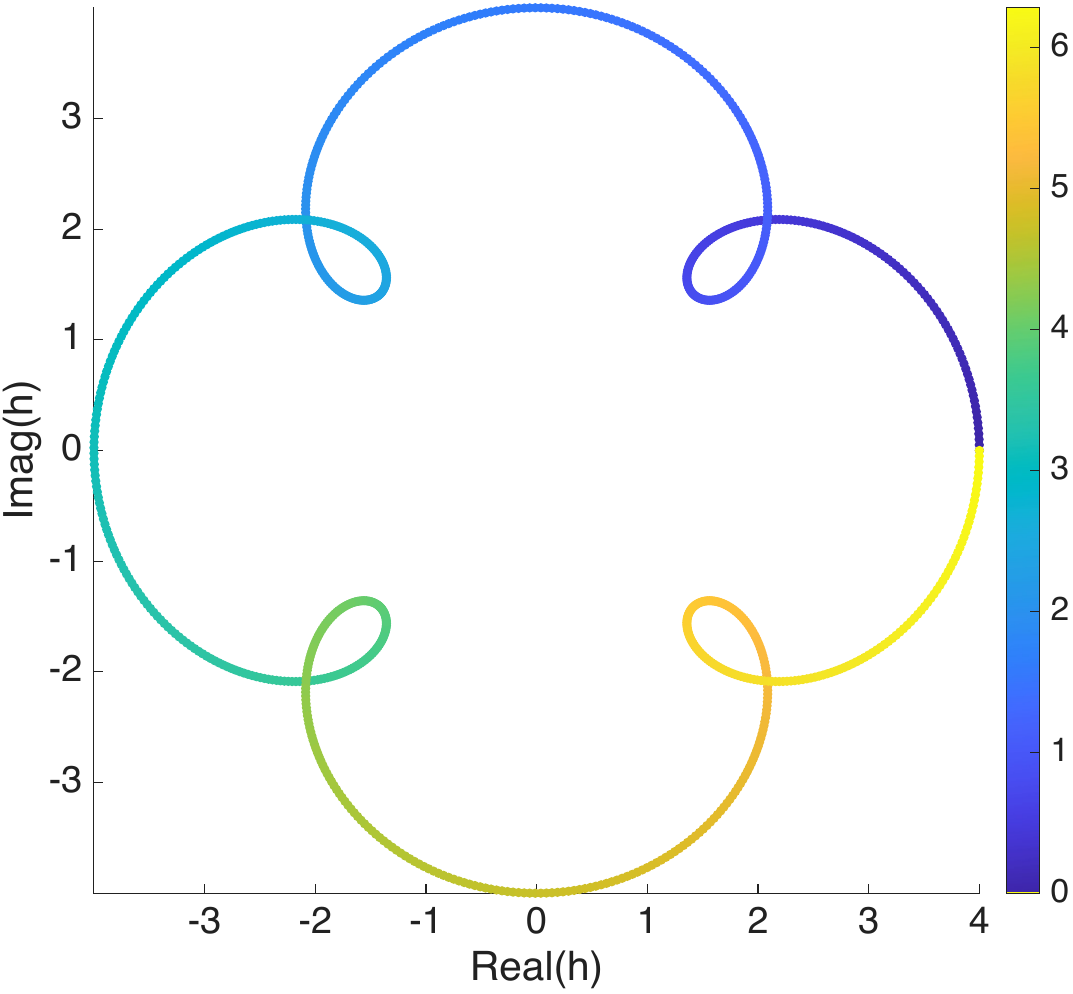}
\caption{The left panel shows $g(e^{i\theta})$ and the right panel shows $h(e^{i\theta})$, where $\theta\in [0,2\pi)$. The colorbar indicates $\theta$. \label{fig winding example}}
\end{figure}

While the PDU framework is theoretically well-founded {(see Section~\ref{section: theoretical results} for a summary)}, several technical challenges arise when applying it to real-world data; {for example, decompose signals satisfying the AHM \eqref{model: AHM equation} into its IMT functions.} We propose an algorithm to address two key limitations of PDU in decomposing signals into their oscillatory components. 
\begin{enumerate}
\item (Challenge 1) PDU relies heavily on the properties of inner functions, a subset of bounded analytic functions ($H^\infty$), which limits the class of AM it can effectively handle for decomposition. While the low-pass filtering strategy discussed in Section \ref{section: review PDU LFP} provides partial relief, the polynomial fitting approach breaks down when the target component of the decomposition exhibits a more complex AM structure that cannot be well approximated by polynomials. See the left panel of Figure \ref{windowBKD} for an example. 

\item (Challenge 2) PDU performs poorly in decomposition when the low-frequency components are weak relative to the high-frequency ones. This has been partially explained by existing result mentioned in the end of Section \ref{section: decomposition and IF results}. Due to PDU's reliance on the signal's winding dynamics, the winding become dominated by high-frequency content when low-frequency components are faint, and results in inaccurate decomposition, which is called {\em mode-mixing}. See the left upper panel of Figure \ref{Nonsinusoidal:2}. This is because the signal winds cross zero erroneously, which leads to a confusion of PDU. Take $g(z)=z+3z^5$ as an example. In this example, there are 5 roots inside $\mathbb{D}$, including $0$ and $e^{i(2j-1)\pi/4}/3^{1/4}$, $j=1,\ldots,4$. Therefore, the first factored component by PDU involves both low frequency component $z$ and high frequency component $5z^3$. On the other hand,  for $h(z)=3z+z^5$, there is only one root $0$ inside $\mathbb{D}$. Thus, the first factored component by PDU is the desired low frequency component $5z$. See Figure \ref{fig winding example} for an example of such phenomenon, where we can easily see that $g$ crosses $0$ various times, while $h$ crosses $0$ only once. Note that while the factored components might not be meaningful, in general the fast convergence of PDU holds. 
See \cite[Proposition 3.2]{Coifman_Steinerberger_Wu:2016} for a related analysis in a simplified polynomial setting.
\end{enumerate}

We now introduce methods to mitigate these issues. These techniques draw on existing insights from related algorithms in the time-frequency (TF) analysis, and their effectiveness will be demonstrated in the following section.

\subsection{Divide-and-conquer by tapering}
{The core idea is a divide-and-conquer approach inspired by TF analysis. In TF analysis, the signal is divided into small pieces, assuming that over each small piece the signal is relatively ``stationary''. The ``spectrum'' of each small piece is then evaluated, and those spectra are patched together to get the TF representation for further analysis. 
This idea motivates a modification of the PDU algorithm to overcome the challenge of globally fitting complex amplitude modulations with polynomials. The modified procedure is composed of the following steps. Suppose the input signal is $f:[0,L]\to \mathbb{C}$, where $L>0$. Note that for any signal recorded over a period $[0,L]$, via a dilation we can assume $L=2\pi$.

\begin{itemize}
\item \textbf{Step 1.}  
Consider a window supported on $[-T,T]$, where $0<T<L$, so that
\begin{equation}\label{window definition w}
    w(t)=\left\{ \begin{array}{lll}
         \sin^2\left(\frac{\pi(t+T)}{2B}\right) & \mathrm{for} & -T\leq t\leq -T+B  \\
         1 & \mathrm{for} & -T+B < t \leq T-B \\
         \cos^2\left(\frac{\pi(t-T+B)}{2B}\right) & \mathrm{for} & T-B < t \leq T
    \end{array} \right. ,
\end{equation}
where $B<T$. Without loss of generality, we choose $T$ and $B$ so that $n:=\frac{L}{2T-B}\in \mathbb{N}$; that is, we will divide the signal into $n$ pieces.

\item \textbf{Step 2.} Divide the input signal $f$ into segments by the chosen window in the following way. Extend $f$ periodically from $[0,L]$ to $[0,L+2T-2B]$ by setting $f(L+t):=f(t)$. For the $j$-th signal piece on $I_j=[(2T-B)(j-1),\,(2T-B)(j-1)+2T]$, $j=1,\ldots,n$, define a new signal over $[0,1)$ via
\begin{equation}\label{windowedPDU truncation dilation}
f_j(t):=f(2Tt+(2T-B)(j-1))w(2Tt-T)\,.
\end{equation}
That is, we truncate the signal over $I_j$ by the window and dilate the truncated signal to be supported on $[0,1)$.

\item \textbf{Step 3.} Apply PDU to each $f_j$ to extract its first PDU-component, denoted by $\tilde f_{j,1}$.

\item \textbf{Step 4.} Stitch together $f_{i,1}$ across all $i$ to form the first decomposed component $\tilde f_1$; that is, set
\[
\tilde{f}_1(t):=\sum_{j\in I_t} f_{j,1}\left(\frac{t-s_j}{e_j-s_j}\right)\,,
\]
where $I_t:=\{j|\, t-s_j\in I_j\}$.

\item \textbf{Iteration.} Repeat Steps 1-4 on $f-\tilde f_1$ to obtain $\tilde f_2$, and continue for subsequent components, $\tilde f_3$, etc.
\end{itemize}
We call this algorithm {\em windowed PDU}.

Some remarks regarding this algorithm are in order.
First, by design, two consecutive windows overlap. If we set $w_i(x)=w(x-i(2T-B))$, then $\sum_{j=1}^nw_j(t)=1$ for all $t\in [B,L+B]$. Consequently, $\sum_{j=1}^nf_j(t)=f(t)$ for $t\in [B,L+B]$. Due to the periodic extension of $f$, this means that the original signal on $[0,T]$ is divided into $n$ segments. 
Second, in practice, the signal is typically nonperiodic, and thus the $n$-th segment $f_n$ is often discontinuous. The resulting component $f_{n,1}$ therefore introduces boundary effects, which are common in many kernel-based algorithms. A simple remedy is to periodize the signal by flipping and stitching it before applying the windowed PDU, which mitigates this discontinuity. 
Third, by construction, the window $w$ is compactly supported and $C^{1,1}$, so each segment $f_l$ is not holomorphic and has at least one zero on the boundary. This situation is not unique to the windowed PDU but is frequently encountered in practice when applying PDU to real world signals. In such cases, a preprocessing step is needed to convert the signal into a holomorphic form, typically by applying the Hilbert transform carefully to avoid generating roots on the boundary, as detailed in Section~\ref{Section: numerical implementation of PDU}.
Fourth, there is flexibility in the choice of window functions. One may use alternative windows $w_j$ with sufficient regularity, and possibly of different lengths, to truncate the signal. Although there is no theoretical guarantee for such choices, empirical evidence suggests that the algorithm performs well provided that the windows are bell-shaped and satisfy $\sum_{j=1}^nw_j(t)=1$.}

{We now take a closer look at how this divide-and-conquer strategy improves the decomposition. Suppose $f(t)$ satisfies the AHM \eqref{model: AHM equation}.
Assume $\epsilon$ is small and each $g_l(t)$ is analytic and the support of $\hat{g}_l$ is away from $0$ (see Section \ref{section: decomposition and IF results} for a discussion about this assumption). 
If $\hat{w}$ is well supported around $0$ and the window length is sufficiently long, $g_l(t)w(t)$ can be well approximated by an analytic function, and the spectral overlap of $g_l(t)w(t)$ and $g_{l'}(t)w(t)$ is negligible. Meanwhile, $T(t)w(t)$, remains a low frequency component. As a result, within each $f_j$,
the trend and the AM of each IMT component can be better approximated by polynomials. Although a complete theoretical analysis of the windowed PDU under the general holomorphic function setup is not yet available, this reasoning explains how Challenge 1 is addressed.

Following this discussion, it is clear that the choice of window width plays a critical role, analogous to the window selection problem in TF analysis.} When approximate prior knowledge of the IFs of different components is available, it can guide the selection of an appropriate window length. Specifically, if the window length $2T$ is shorter than the period of the low frequency component, e.g., $1/\phi'_1(t)$, the low frequency component in $f_j$ behaves like a trend since it does not complete a full oscillation. This enables the high-frequency component to be extracted first. Once the high-frequency components are identified, one can successively enlarge the window to extract lower-frequency components. This strategy helps avoid mode mixing and provides a practical solution to Challenge 2. Such prior information is often available in biomedical signals. For instance, in the PPG signals we study later, it is generally known that they contain cardiac and respiratory components, whose typical frequency ranges are well established in the literature except in rare pathological cases. A systematic and automatic approach toward such width choice will be explored in our future work.

\begin{figure}[hbt!]
\includegraphics[trim=0 0 0 0, clip,width=1\textwidth]{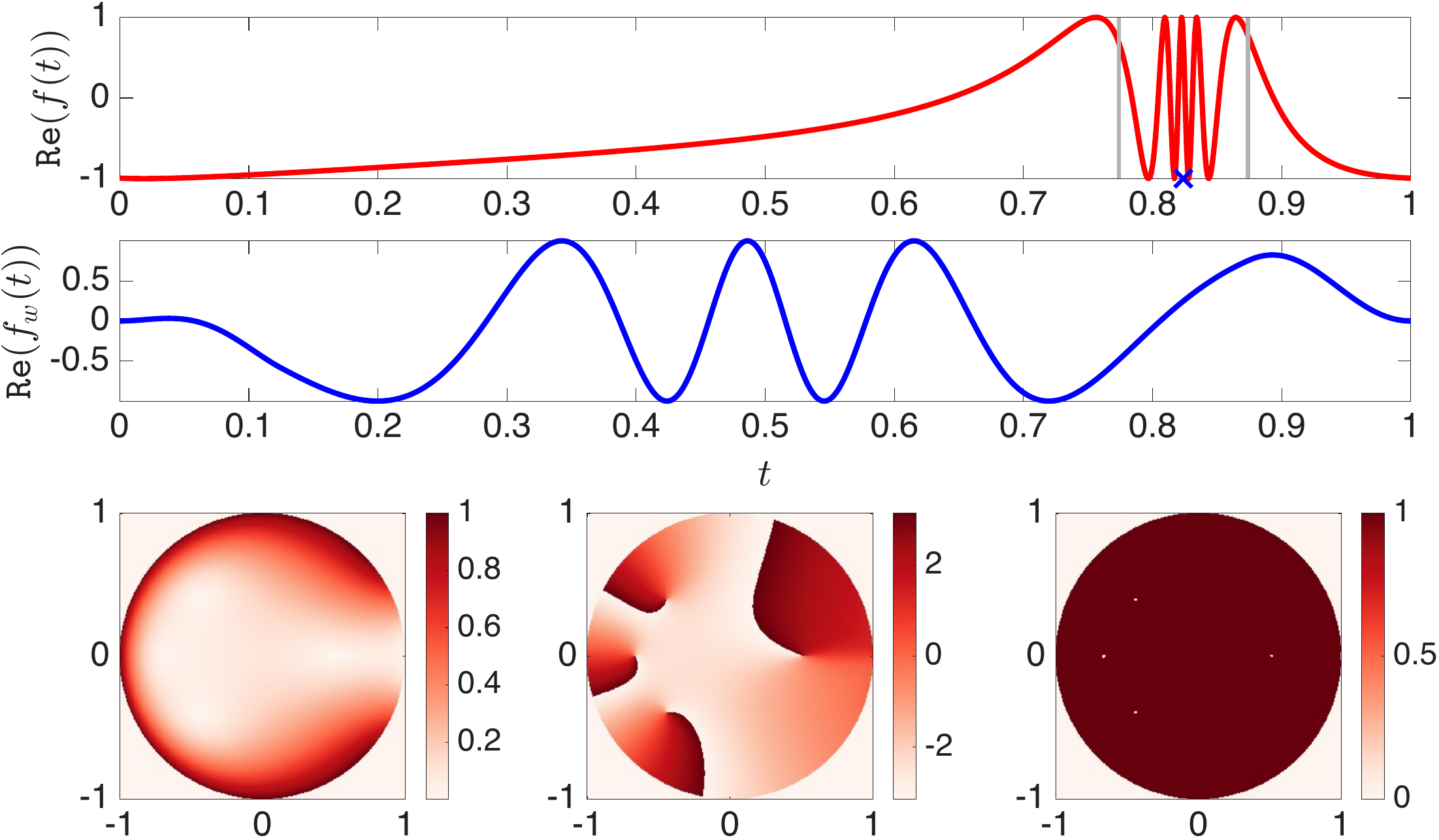}
\caption{Illustration of the window effect in the proposed windowed PDU algorithm. Consider $f(t)=f_\alpha^5(t)$, where $f_\alpha(t):=\frac{e^{i2\pi t}-\alpha}{1-\overline{\alpha}e^{i2\pi t}}$, $t\in [0, 1)$ and $\alpha=\frac{9-18i}{10\sqrt{5}}$. The top panel shows the real part of $f$, with the gray vertical lines indicates the truncation range. The window is defined by $B=0.045$ and $T=B/4$. The resulting truncated signal, denoted as $f_w$, is shown in the middle panel. In the bottom left panel, the magnitude of the holomorphic function $F_w=P_+f_w$ over $\mathbb{D}$ evaluated by Poisson integration, is displayed. In the bottom middle panel, the phase of the holomorphic function $F_w$ over $\mathbb{D}$ is displayed. In the bottom right panel, the region where $|F_w|<0.005$ is shown in white, while all other regions are shown in red.   \label{fig windowed PDU example}}
\end{figure}

The windowed PDU algorithm is directly linked to the relationship between roots and IF. In Figure \ref{fig Blaschke product example}, we illustrate how an oscillatory template becomes temporally localized depending on the position of its associated root. For instance, consider the function $f_{\alpha_3}^4$ in Figure \ref{fig Blaschke product example}. A sufficiently wide window truncation centered around $0.82$ captures the entire oscillation pattern. The truncation and dilation to the interval $[0,1)$ in \eqref{windowedPDU truncation dilation} numerically corresponds to shifting the root near the boundary towards zero. 
See Figure \ref{fig windowed PDU example} for a numerical example of this intuitive description, where we consider $f(t)=f_\alpha^5(t)$, $\alpha=\frac{9-18i}{10\sqrt{5}}$,   $f_\alpha(t):=\frac{e^{i2\pi t}-\alpha}{1-\overline{\alpha}e^{i2\pi t}}=e^{i\phi_\alpha(t)}$, and $t\in [0, 1)$. The window is described in \eqref{window definition w} and centered at 0.82 with $B=0.045$ and $T=B/8$. 
The resulting truncated and dilated signal is denoted as $f_w$. The holomorphic function of $P_+f_{j}$ over $\mathbb{D}$ is then evaluated by the Poisson integration, denoted by $F_w$. In the bottom right panel, the region with $|F_w|$ less than $0.005$ is shown in white, while all others are shown in red. All potential roots have magnitude less than 0.7. It suggests that while the original root is close to the boundary, after truncation, the roots are closer to zero, {reflecting that the oscillation varies less over time following truncation and dilation.}
Intuitively, this operation resembles a \emph{local} version of M\"obius transform, if properly defined. We hypothesize that developing a rigorous quantification of this concept may provide a deeper understanding of the underlying mechanism of the windowed PDU algorithm.

\begin{figure}[hbt!]
\includegraphics[trim=0 0 0 0, clip,width=1\textwidth]{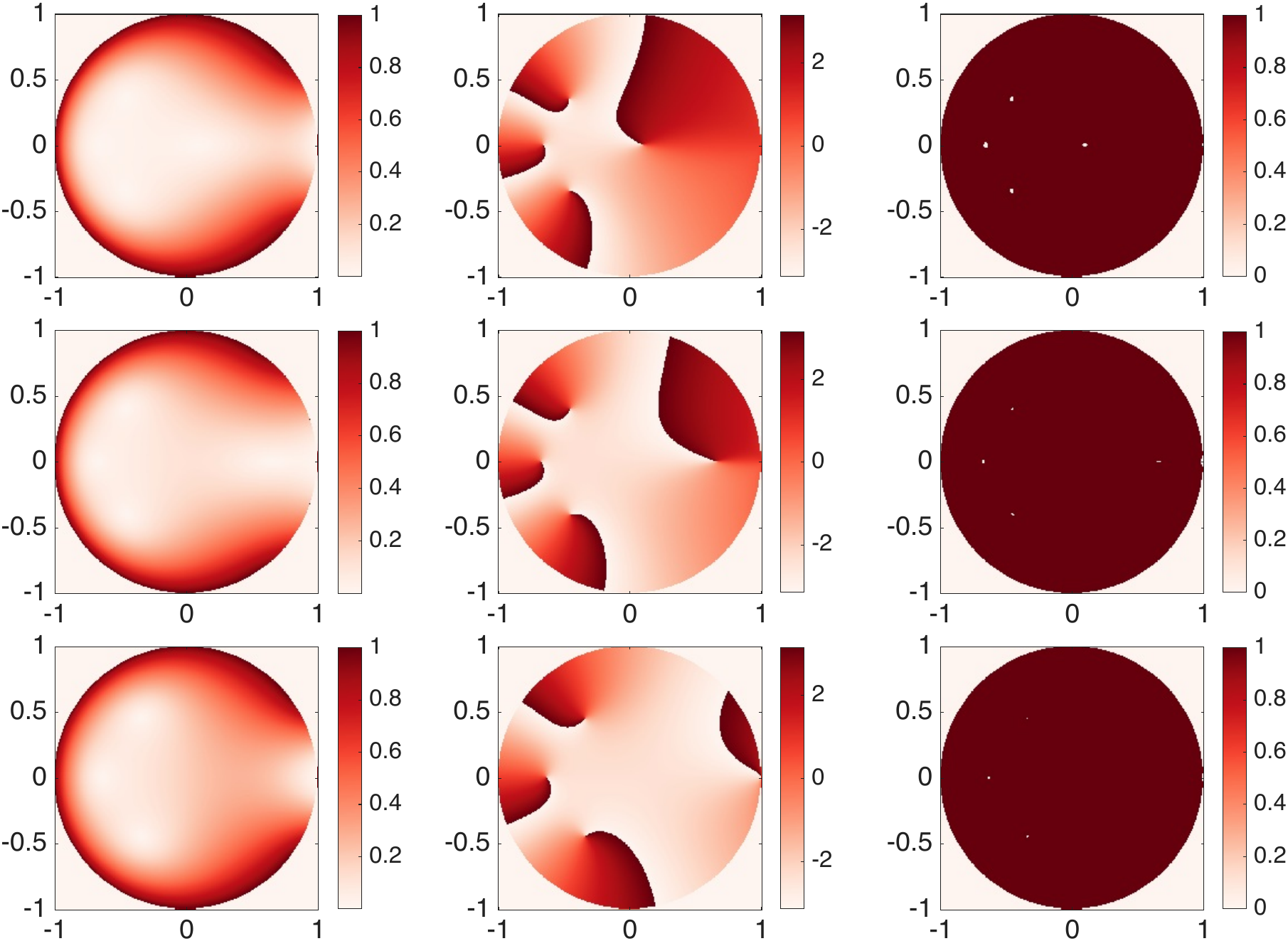}
\caption{A continuous of Figure \ref{fig windowed PDU example} with the same signal and meaning of each panel, while from top to bottom rows the window described in \eqref{window definition w} are centered at 0.82 with $B=0.055$ and $T=B/8$, $B=0.045$ and $T=B/6$ and $B=0.035$ and $T=B/9$ respectively.   \label{fig windowed PDU comparison}}
\end{figure}

Another numerical observation is that the roots, particularly those whose phases are far from $0$ (associated with the tapering region of the window), exhibit ``stability'' with respect to the choice of window. See Figure \ref{fig windowed PDU comparison} for an example. From top to bottom, the window is defined in \eqref{window definition w} and centered at 0.82, with parameters $B=0.055$ and $T=B/8$, $B=0.045$ and $T=B/6$ and $B=0.035$ and $T=B/9$, respectively. 
Clearly, although the phase plots vary across different windows, the locations of roots with phases far from 0 are less sensitive to the choice of window. Since these roots capture the primary oscillatory information, this recalls the ``spin cycling'' technique for denoising \cite{coifman1995translation}. Rather than relying on a single window, one could instead explore multiple windows and design an ``averaging'' strategy that assigns greater weight to the more stable roots. We hypothesize that this approach, analogous to spin cycling, could enhance the decomposition and potentially improve denoising. We plan to investigate this direction in future work.

\subsection{Anti-derivative as a solution}

A practical challenge when applying PDU to analyze biomedical signals is when the low frequency component might be weaker than the high frequency component, which is listed above as Challenge 2. In addition to the tapering idea, another approach to resolve Challenge 2 is the following cumsum idea.

First, apply cumsum to the input signal $f$, and obtain a new function $\bar{f}$. To enhance the accuracy of cumsum, we recommend to upsample the signal and remove the polynomial trend from $\bar{f}$ by fitting a quadratic or cubic polynomial to $\bar{f}$. Apply PDU or windowed PDU to $\bar{f}$, and denote the obtained decomposed components as $\bar{f}_l$, where $l=1,2,\ldots$. The final decomposed components are denoted as $\tilde{f}_l:=\bar{f}_l'$, the derivative of $\bar{f}_l$.

The basic idea beyond this proposal is the ``low pass'' filter idea of antiderivative. Note that if we have $f(t)=\sum_{l=1}^n\alpha_l\cos(2\pi\xi_lt)$, where $\alpha_l>0$ and $0<\xi_1<\xi_2<\ldots$, by the fundamental theory of calculus, we obtain $\bar{f}(t)=\sum_{l=1}^n\frac{\alpha_l}{\xi_l}\sin(2\pi\xi_lt)+c$ for some $c\in \mathbb{R}$. 
In other words, antiderivative enhances the strength of low-frequency harmonic function and suppress the strength of high-frequency harmonic functions, so that the weaker low frequency component problem in Challenge 2 is alleviated.

\section{Numerical example}\label{sec: simu and real data}

The Matlab codes of PDU and the proposed windowed PDU used to generate the following results can be found in {\url{https://github.com/hautiengwu2/Phase-Dynamics-Unwinding.git}} for the reproducibility purposes. 

\subsection{Simulated signal}

We simulate AHM signals in the following way. Denote the standard Gaussian random walk on $\mathbb{Z}$ as $W$ so that $W(0)=0$. Smoothen $W$ with the local regression using weighted linear least squares and a second degree polynomial model, where the number of data points for calculating the smoothed value is denoted as $B\in \mathbb{N}$. Denote the result $W_B$.
Denote the sampling rate as $f_s>0$. Over $[0,T_0]$, set the random vectors of dimension $\lfloor f_sT_0\rfloor$ so that
\[
A_1(l)=a_1\left(1+\frac{\alpha_1|W_{2.2f_s}(l)|}{\max_j|W_{2.2f_s}(j)|}\right)\,,\ 
A_2(l)=a_2\left(1+\frac{\alpha_2|W_{2f_s}(l)|}{\max_j|W_{2f_s}(j)|}\right)\,,
\]
\[
\phi'_1(l)=\xi_1+\frac{\beta_1|W_{2.2f_s}(l)|}{\max_j|W_{2.2f_s}(j)|}\,,\mbox{ and }
\phi'_2(l)=\xi_2+\frac{\beta_2|W_{2f_s}(l)|}{\max_j|W_{2f_s}(j)|}\,,
\]
where $l=1,\ldots,\lfloor f_sT_0\rfloor$, $a_1$, $a_2$, $\alpha_1$, $\alpha_2$, $\xi_1$, $\xi_2$, $\beta_1$, and $\beta_2\geq0$ are set differently.
Then, independently realize $A_1$, $A_2$, $\phi'_1$ and $\phi'_2$ so that $\min(\phi'_2(l)-\phi'_1(l))>0.5$; that is, force IFs of both IMT functions to be separated by $0.5$ Hz. Set $\phi_1,\phi_2\in \mathbb{R}^{\lfloor f_sT_0\rfloor}$ by 
\[
\phi_k(l)=\sum_{j=1}^l \phi'_k(j)/f_s\,, 
\]
where $k=1,2$.
The final signal is $f\in \mathbb{R}^{\lfloor f_sT_0\rfloor}$ so that 
\[
f(l)=A_1(l)\cos(2\pi\phi_l(l))+A_2(l)\cos(2\pi\phi_2(l))\,. 
\]
This signal has two IMT functions, and the IMT function with lower frequency has a larger amplitude. 
Below, set $f_s=512$ and $T_0=16$. When we run PDU, we always upsample the signal by 16 times, and use $L=5$ in \eqref{PDUexpansion2}. In the windowed PDU, the tapering window $w$ in \eqref{window definition w} is set with $T=1/4$ and $B=T/4$.

To quantify the performance of each proposed technique, we consider the following indices. Denote $\tilde{f}_l\in \mathbb{C}^{\lfloor f_sT\rfloor}$ to be the $l$-th decomposed component by the PDU algorithm. The first index is the normalized root mean square error (NRMSE) of AM recovery by
\[
\Delta^{(1)}_l := \frac{\||\tilde{f}_l|-A_l\|_2}{\|A_l\|_2}\,,
\]
where $l=1,2$.  The second index is the standard deviation (SD) of the phase recovery by
\[
\Delta^{(2)}_l:=\texttt{SD}\{\angle(\tilde{p}_l(k)e^{-i\phi_l(k)})|\, k=1,\ldots,\lfloor f_sT\rfloor\}\,,
\]
where $\tilde{p}_l\in \mathbb{C}^{\lfloor f_sT\rfloor}$ so that $\tilde{p}_l(j)=\tilde{f}_l(j)/|\tilde{f}_l(j)|$ and $l=1,2$. The third one is the NRMSE of the recovery of the signal by
\[
\Delta^{(3)}:=\frac{\|f-\Re(\tilde{f}_1+\tilde{f}_2)\|_2}{\|f\|_2}\,.
\]

\begin{figure}[hbt!]
\includegraphics[trim=0 0 0 0, clip,width=0.495\textwidth]{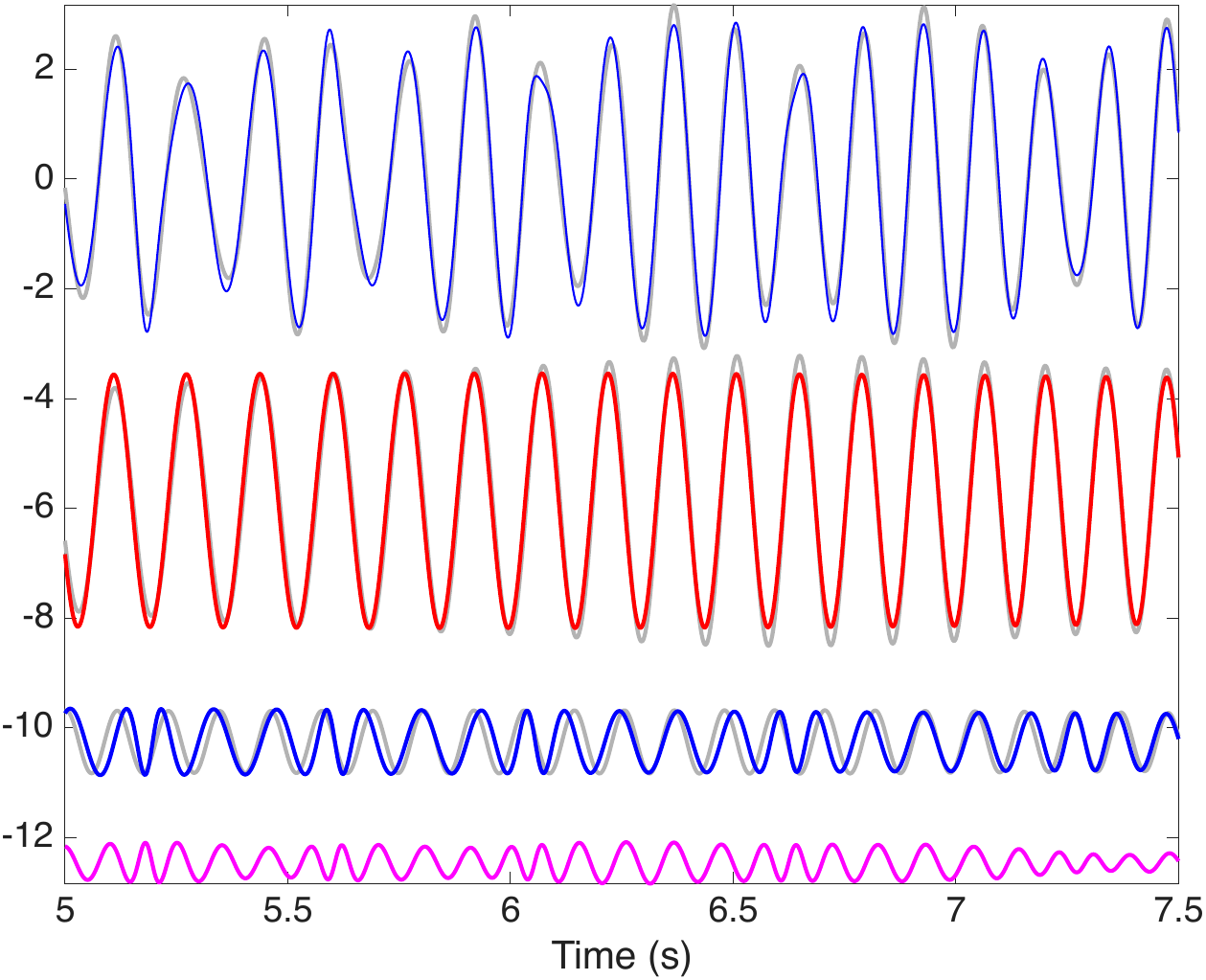}
\includegraphics[trim=0 0 0 0, clip,width=0.495\textwidth]{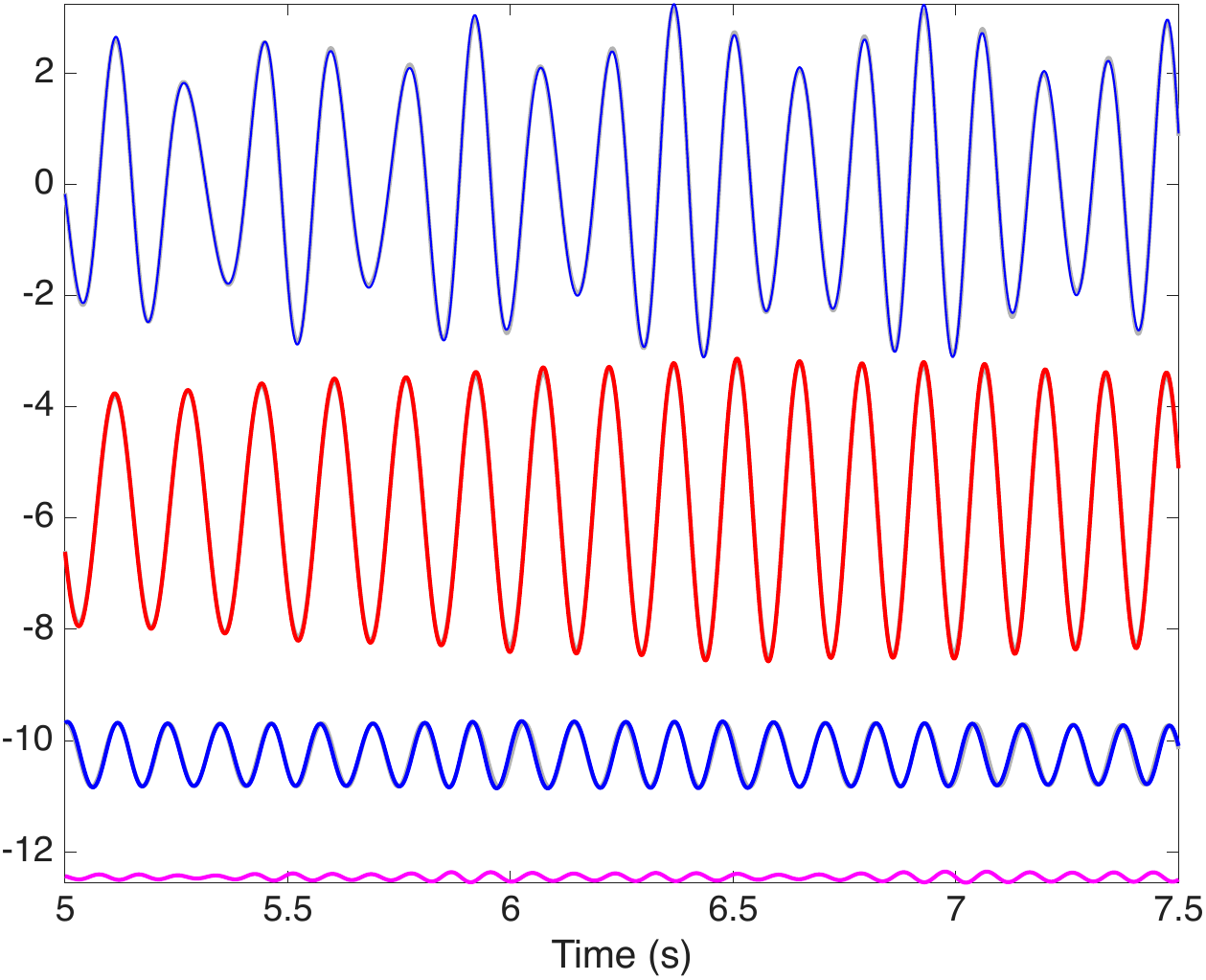}
\caption{Left panel: results of PDU; right panel: results of windowed PDU. In each panel, the gray curve shown in the top is the input signal $f\in \mathbb{R}^{\lfloor f_sT\rfloor}$, which is the summation of two IMT functions shown as gray curves below. The decomposed components by PDU and windowed PDU are superimposed as red and blue curved, and the summation of the decomposed components, $\Re(\tilde{f}_1+\tilde{f}_2)$, is superimposed on the top black curve. The difference between $f-\Re(\tilde{f}_1+\tilde{f}_2)$ is shown in the bottom magenta curve. To enhance the visualization, only the middle 2.5 seconds are shown. \label{windowBKD}}
\end{figure}

\begin{figure}[hbt!]
\includegraphics[trim=0 0 0 0, clip,width=0.995\textwidth]{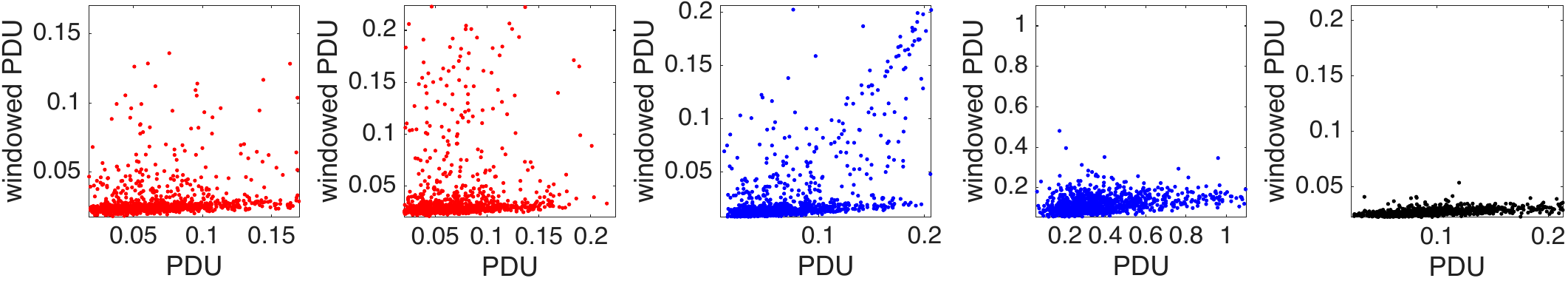}
\caption{Comparison of PDU and windowed PDU under the setup used in Figure \ref{windowBKD}. From left to right, we show the scatterplots of $\Delta^{(1)}_1$, $\Delta^{(1)}_2$, $\Delta^{(2)}_1$, $\Delta^{(2)}_2$, and $\Delta^{(3)}$ over $1000$ realizations of clean signals.  \label{windowBKDsummary}}
\end{figure}

See Figure \ref{windowBKD} for an example comparing PDU and windowed PDU when the low-frequency IMT has a larger AM and noise does not exist, where we set $a_1=2$, $a_2=0.8$, $\alpha_1=\alpha_2=1$, $\xi_1=2+\pi$, $\xi_2=8$, $\beta_1=2.5$ and $\beta_2=3$. It is clear that windowed PDU better reconstructs each IMT function compared with the original PDU. Particularly, the AM of the low-frequency IMT function and the phase of the high-frequency IMT function are better recovered by windowed PDU. In this example, 
\[
(\Delta^{(1)}_1, \Delta^{(1)}_2, \Delta^{(2)}_1, \Delta^{(2)}_2,\Delta^{(3)})=(0.103\,, 0.039 \,,   0.101\,,   0.950 \,,   0.140)
\]
when applying PDU, and 
\[
(\Delta^{(1)}_1, \Delta^{(1)}_2, \Delta^{(2)}_1, \Delta^{(2)}_2,\Delta^{(3)})=(0.027\,,   0.033 \,,  0.013 \,,   0.172 \,,   0.029)
\]
when applying windowed PDU. This obviously shows the improvement of both PDU and windowed PDU.  If we realize $f\in \mathbb{R}^{\lfloor f_sT\rfloor}$ independently for $1000$ times, the overall behavior of PDU and windowed PDU is shown in Figure \ref{windowBKDsummary}. Overall, the performance of windowed PDU is better than that of PDU in all indices with statistical significance if we apply the Wilcoxon signed-rank test with the Bonferroni correction and view $p<0.05$ as statistical significance.  

\begin{figure}[hbt!]
\includegraphics[trim=0 0 0 0, clip,width=0.495\textwidth]{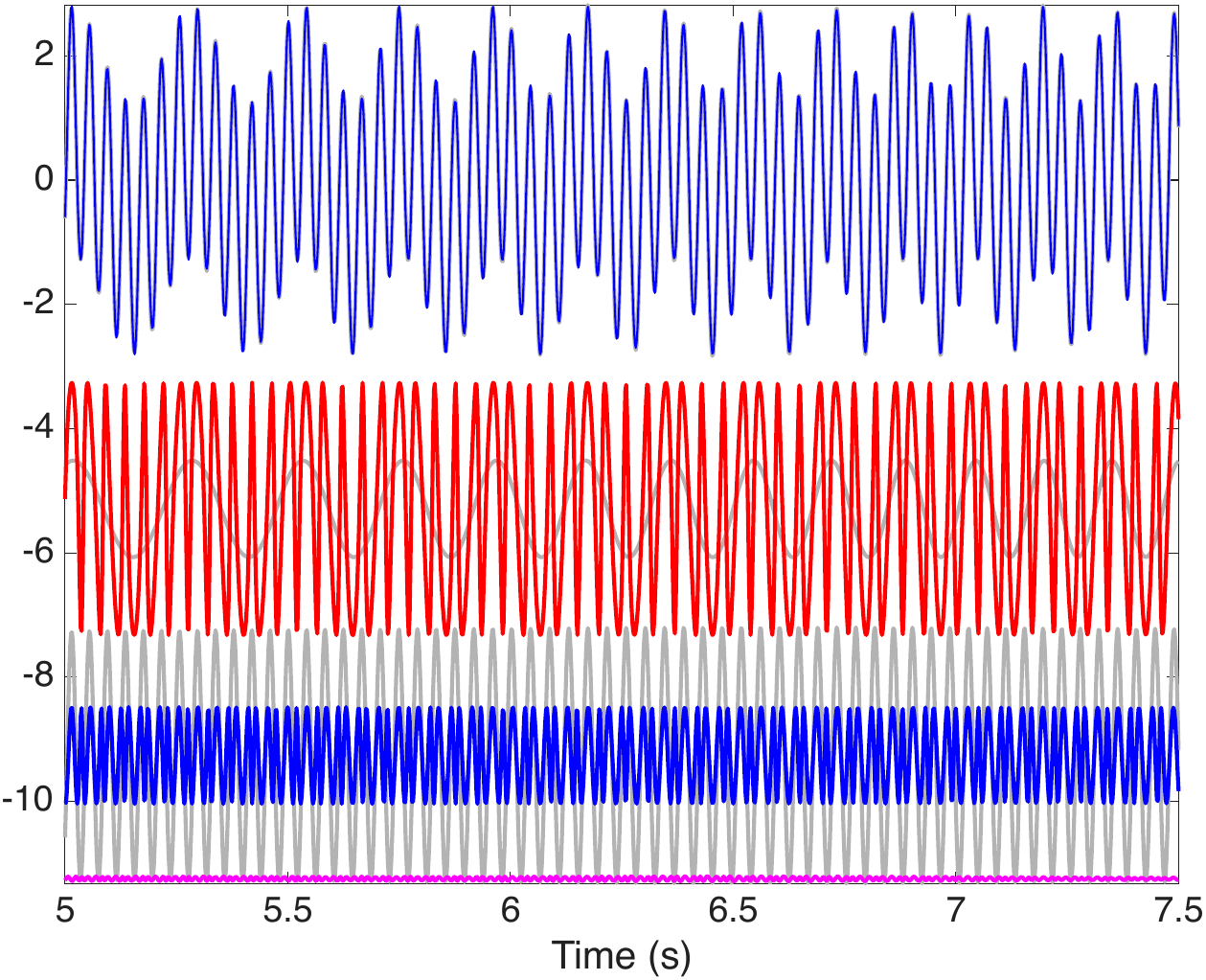}
\includegraphics[trim=0 0 0 0, clip,width=0.495\textwidth]{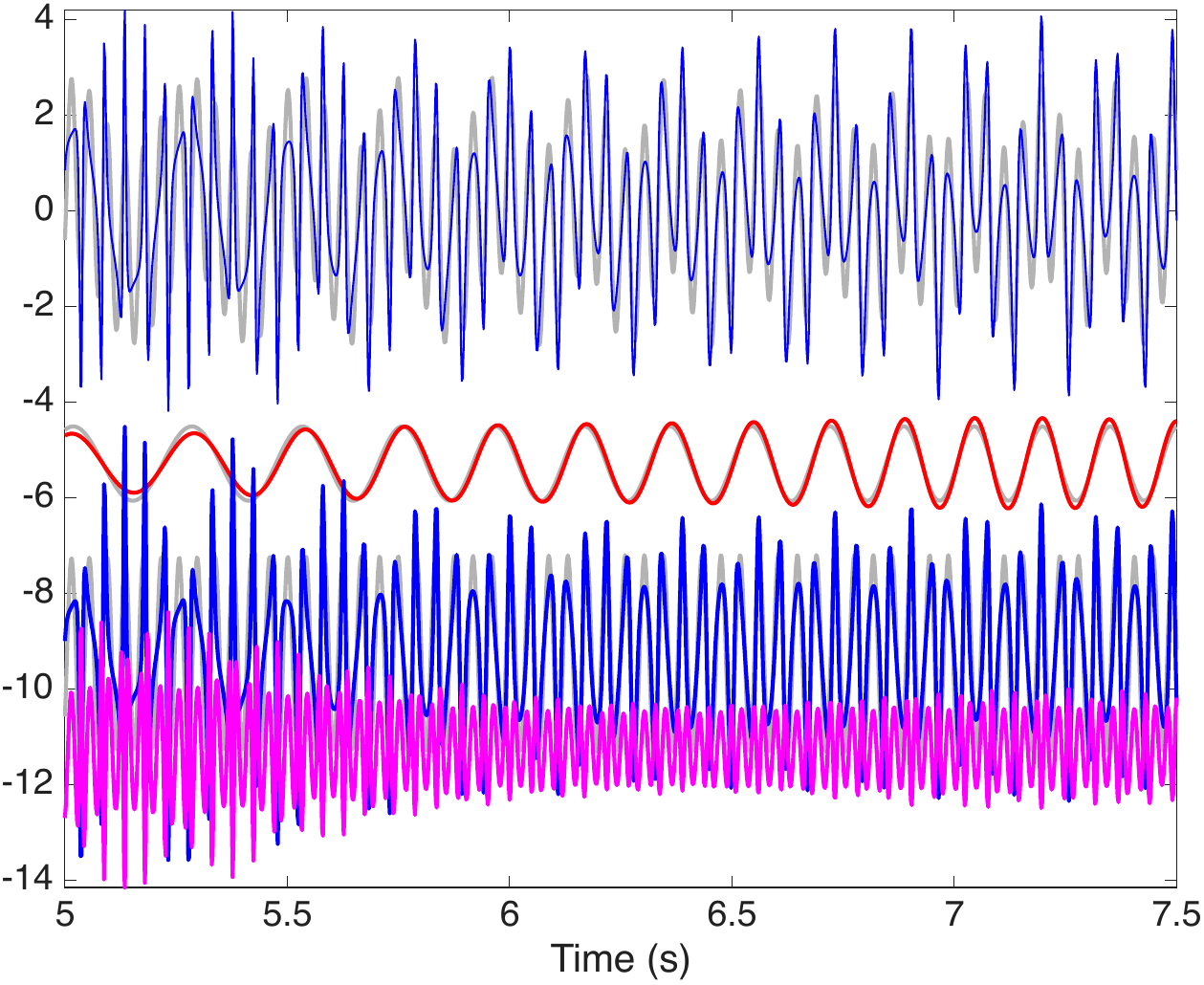}
\includegraphics[trim=0 0 0 0, clip,width=0.495\textwidth]{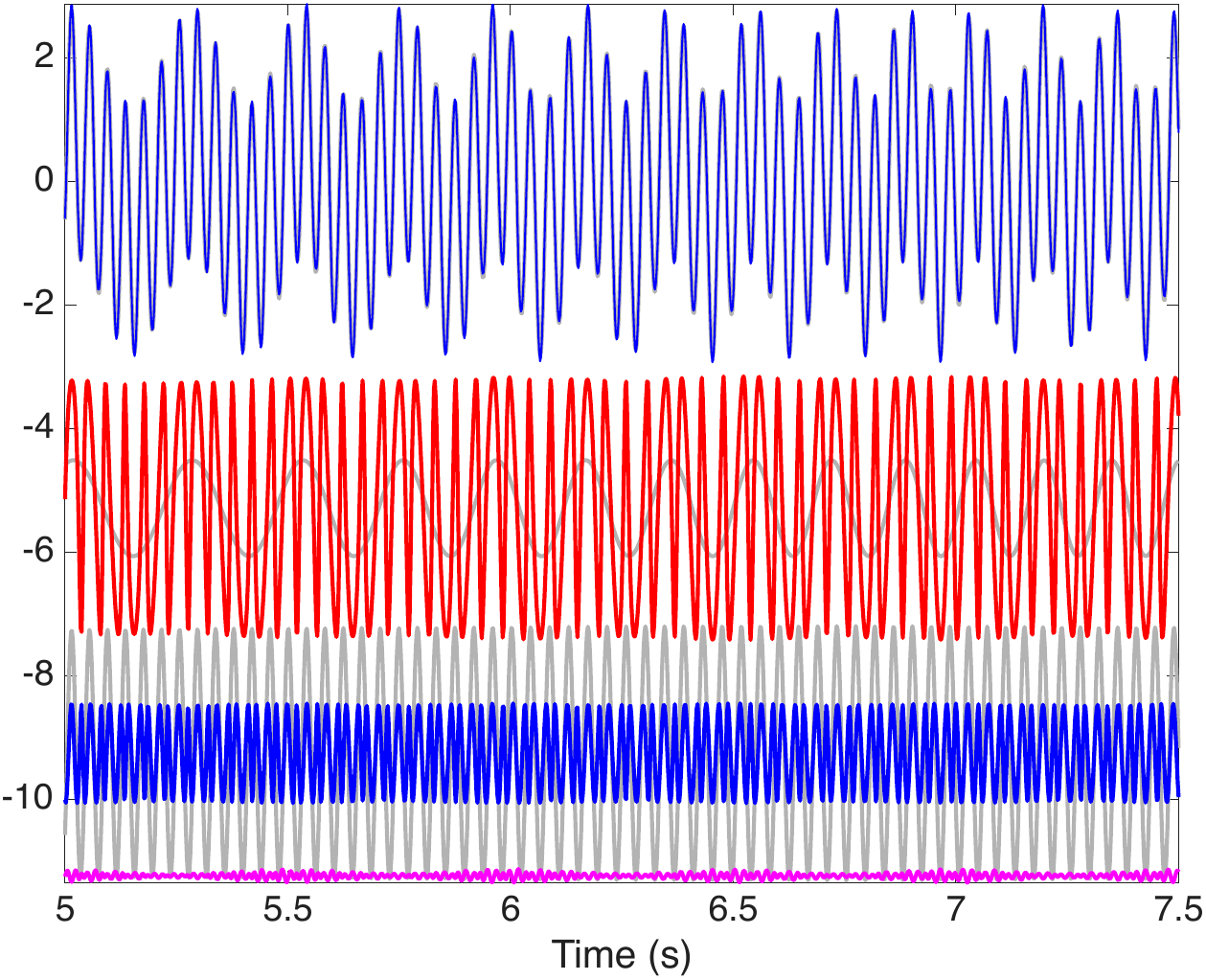}
\includegraphics[trim=0 0 0 0, clip,width=0.495\textwidth]{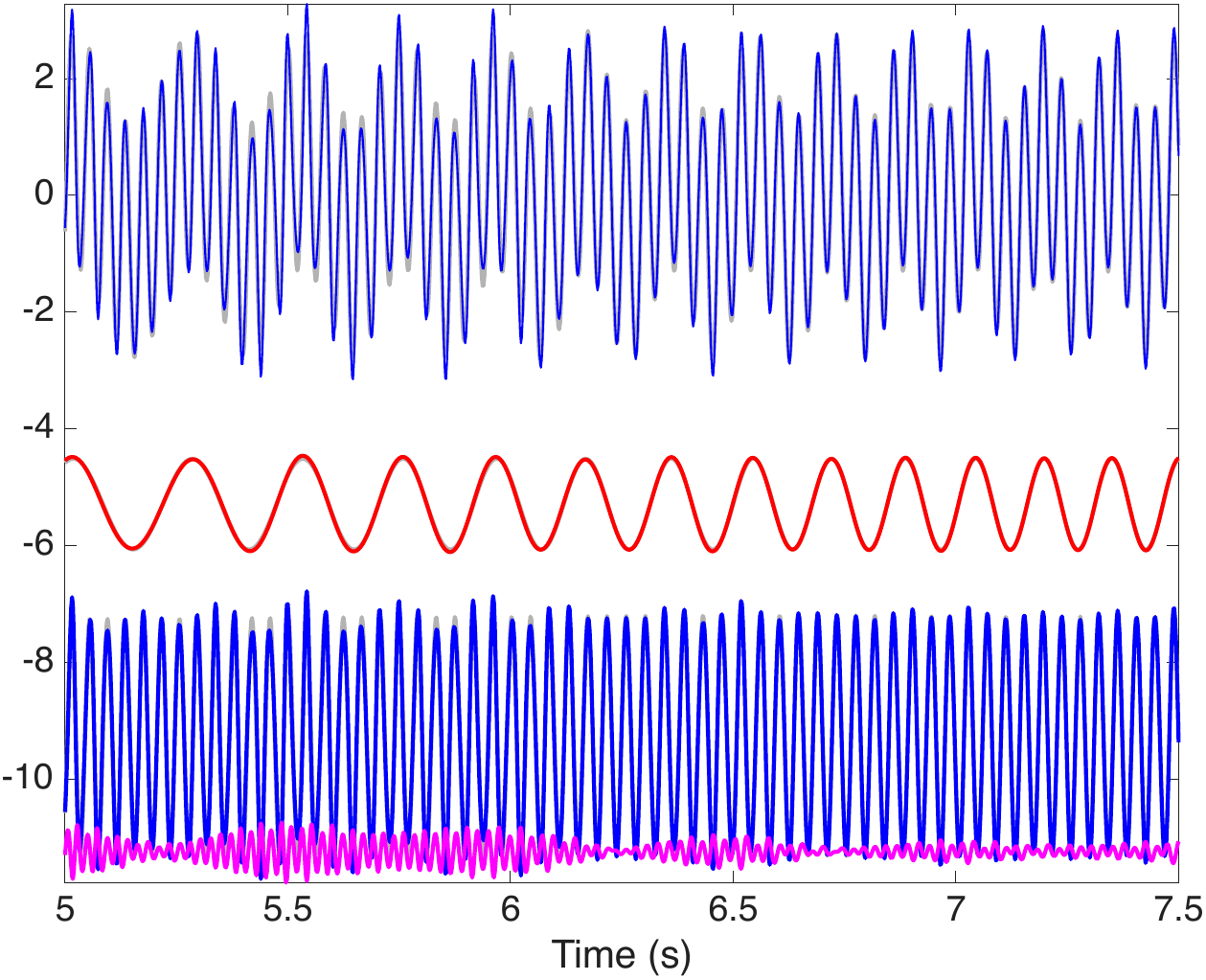}
\caption{Top panels: results of PDU; bottom panels: results of windowed PDU; left panels: without applying the cumsum technique; right panels: with the cumsum technique applied. In each panel, the gray curve shown in the top is the input signal $f\in \mathbb{R}^{\lfloor f_sT\rfloor}$, which is the summation of two IMT functions shown as gray curves below. The decomposed components by PDU and windowed PDU are superimposed as red and blue curved, and the summation of the decomposed components, $\Re(\tilde{f}_1+\tilde{f}_2)$, is superimposed on the top black curve. The difference between $f-\Re(\tilde{f}_1+\tilde{f}_2)$ is shown in the bottom magenta curve. To enhance the. visualization, only the middle 2.5 seconds are shown.
\label{Nonsinusoidal:2}}
\end{figure}

See Figure \ref{Nonsinusoidal:2} for an example comparing PDU and windowed PDU when the low-frequency IMT has a smaller AM and noise does not exist, where we set $a_1=0.8$, $a_2=2$, $\alpha_1=\alpha_2=0.1$, $\xi_1=2+\pi$, $\xi_2=26$, $\beta_1=2.5$ and $\beta_2=3$. In this case, PDU and windowed PDU both fail. However, with the cumsum technique, both PDU and windowed PDU succeed in decomposing both components, and windowed PDU better reconstructs each IMT function compared with the original PDU. In this example, if cumsum is not applied, we have  $$(\Delta^{(1)}_1, \Delta^{(1)}_2, \Delta^{(2)}_1, \Delta^{(2)}_2,\Delta^{(3)})=(1.611\,, 0.617 \,,   1.374\,,  1.374 \,,   0.016)$$ when applying PDU, and $$(\Delta^{(1)}_1, \Delta^{(1)}_2, \Delta^{(2)}_1, \Delta^{(2)}_2,\Delta^{(3)})=(1.677\,,   0.607 \,,  1.374 \,,  1.433 \,,   0.028)$$ when applying windowed PDU. Clearly, both PDU and windowed PDU fail.
After applying cumsum, we have
$$(\Delta^{(1)}_1, \Delta^{(1)}_2, \Delta^{(2)}_1, \Delta^{(2)}_2,\Delta^{(3)})=(0.099\,, 0.257 \,,   0.099\,,   0.326 \,,   0.380)$$
when applying PDU, and $$(\Delta^{(1)}_1, \Delta^{(1)}_2, \Delta^{(2)}_1, \Delta^{(2)}_2,\Delta^{(3)})=(0.028\,,   0.060 \,,  0.017 \,,   0.075 \,,   0.090)$$ when applying windowed PDU. 
Note that compared with Figure \ref{windowBKD}, the high frequency component in this setup has a much higher frequency, and the variation of AM is smaller. This is designed intentionally to guarantee that after cumsum, the low frequency component has a dominant amplitude. 

\begin{figure}[hbt!]
\includegraphics[trim=0 0 0 0, clip,width=0.995\textwidth]{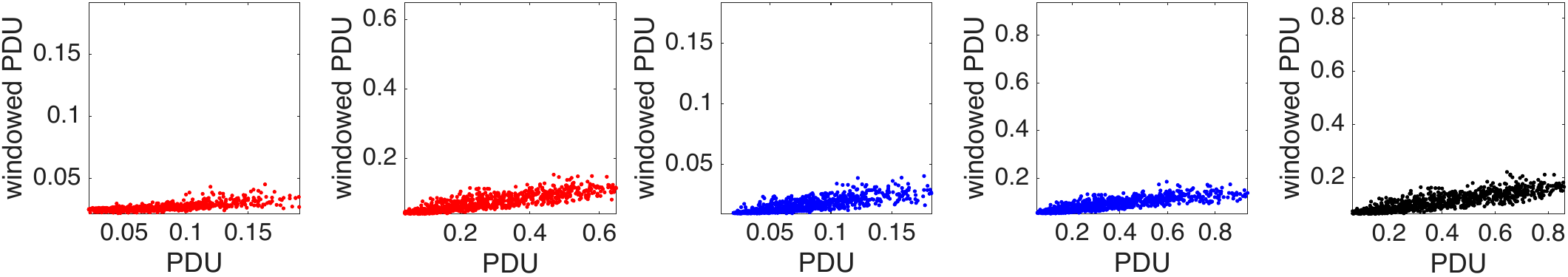}
\caption{Comparison of PDU and windowed PDU under the setup used in Figure \ref{Nonsinusoidal:2}. From left to right, we show the scatterplots of $\Delta^{(1)}_1$, $\Delta^{(1)}_2$, $\Delta^{(2)}_1$, $\Delta^{(2)}_2$, and $\Delta^{(3)}$ over $1000$ realizations of clean signals.  \label{Nonsinusoidal:2summary}}
\end{figure}

Visually, we see that with the cumsum technique, $\Delta^{(3)}$ becomes larger in both PDU and windowed PDU. This larger error comes from the numerical implementation of cumsum and derivative.  
If we realize $f\in \mathbb{R}^{\lfloor f_sT\rfloor}$ independently for $1000$ times, the overall behavior of PDU and windowed PDU is shown in Figure \ref{Nonsinusoidal:2summary}. Overall, the performance of windowed PDU is better than that of PDU in all indices with statistical significance if we apply the Wilcoxon signed-rank test with the Bonferroni correction and view $p<0.05$ as statistical significance.  

\subsection{Real data}

Consider a photoplethysmogram (PPG) signal sampled at 100 Hz. Physiologically, PPG captures not only hemodynamic (cardiac) activity but also respiratory dynamics \cite{shelley2007photoplethysmography} through a phenomenon known as respiratory-induced intensity variation (RIIV) \cite{johansson1999estimation}. RIIV modulates the baseline and amplitude of the PPG waveform in synchrony with respiration, providing a noninvasive means to extract respiratory information.
Mathematically, when RIIV is present, the PPG signal consists of two dominant oscillatory components: a high-frequency cardiac component and a low-frequency respiratory component. Typically, the respiratory component exhibits a smaller amplitude compared to the cardiac component.
This imbalanced amplitude poses a challenge for the PDU algorithm, listed above as Challenge 2. 

\begin{figure}[hbt!]
    \centering
\includegraphics[trim=0 0 0 0, clip, width=\columnwidth]{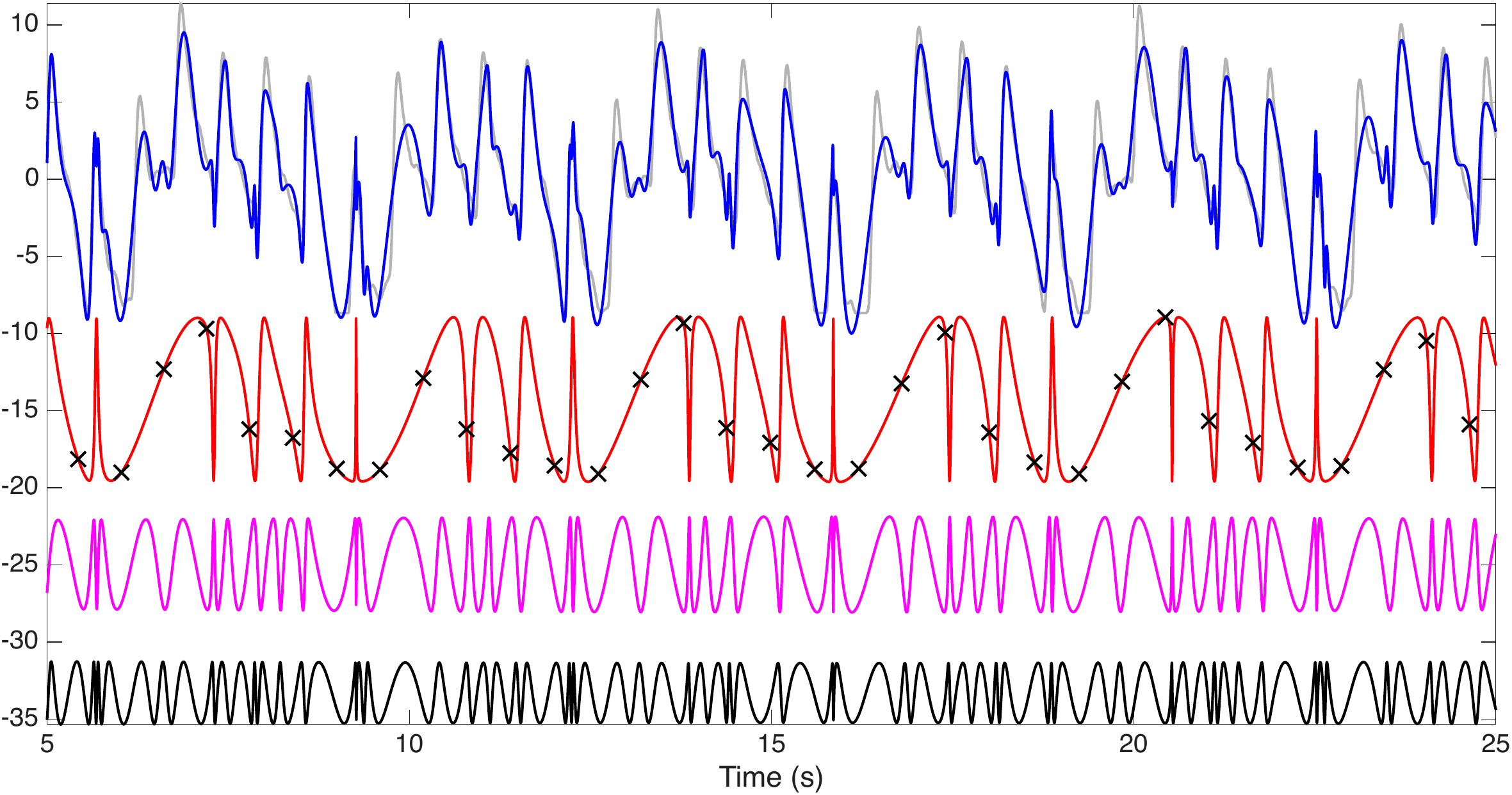}
\includegraphics[trim=0 0 0 0, clip, width=\columnwidth]{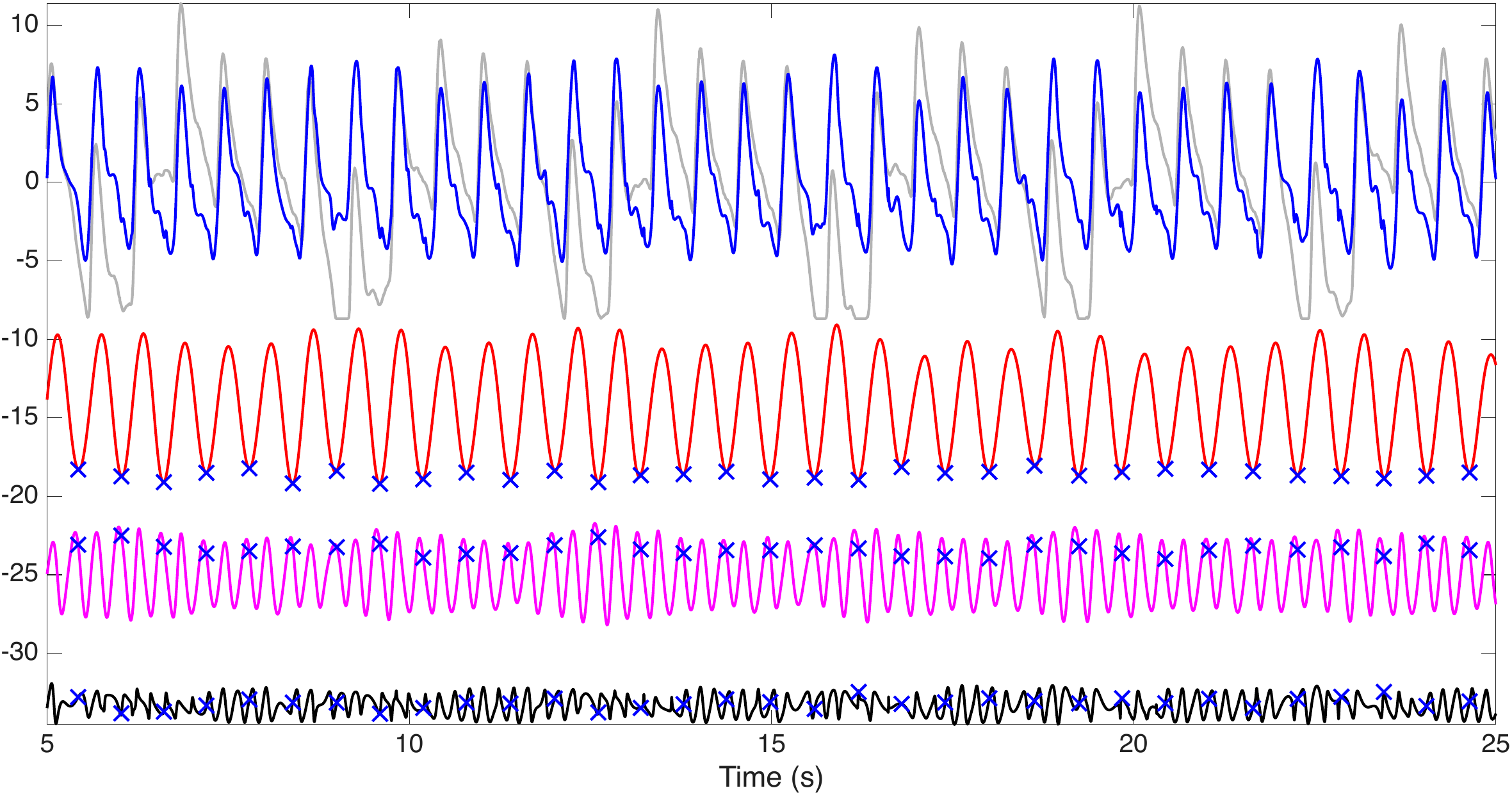}
    \caption{Comparison of PDU and windowed PDU on a PPG signal. The top panel shows the result of PDU, and the bottom panel shows the result of windowed PDU. The input PPG signal is plotted in gray. The first three decomposed components are shown in red, magenta, and black (from bottom to top), and their sum is superimposed on the input signal in blue. Black crosses indicate the timing of R-peaks from the simultaneously recorded electrocardiogram signal.}
    \label{fig illustration windowPDU vs PDU PPG}
\end{figure}

The results of the windowed PDU and its comparison with the PDU are shown in Figure \ref{fig illustration windowPDU vs PDU PPG}.  In the top panel, the first three components extracted by PDU using a fifth-order low-pass filter are shown in different colors, and their sum is overlaid on the original PPG signal as a blue curve. It is evident that the cardiac and respiratory components are not properly separated in the PDU decomposition. This mode-mixing hampers physiologically meaningful interpretation of each individual component. Nonetheless, the overall PPG signal is well-approximated by the sum of just three components  with the normalized root mean squared error (NRMSE) 0.300. 
In the bottom panel, the top three decomposed components obtained using the windowed PDU with the window function $w$ defined in \eqref{window definition w} and $T=1$ and $B=1/2$ are shown in different colors, with their sum superimposed as a blue curve. Here, we use the physiological knowledge to determine $T$ so that the window length is chosen to be shorter than the respiratory period but long enough to capture the cardiac oscillation. As a result, we expect the cardiac component to be effectively decomposed, and this expectation is met visually.

To better understand this result and justify the successful extraction of the cardiac component, recall that the AHM \eqref{model: AHM equation} can be generalized to accommodate non-sinusoidal waveforms by replacing $\cos(2\pi t)$ with a $1$-periodic function $s$ that captures the morphology of each oscillatory component; that is, the PPG signal can be modeled as $f(t)=A_r(t)s_r(\phi_r(t))+A_c(t)s_c(\phi_c(t))$, where the subscripts $r$ and $c$ refer to the respiratory and cardiac components, respectively, and $s_r,\, s_c$ are  the so-called wave-shape functions that describe their morphologies \cite{Wu:2013}. Using a Fourier series expansion, this model becomes
\[
f(t)=\sum_{k=1}^\infty A_r(t)\alpha_{r,k}\cos(2\pi k\phi_r(t)+\beta_{r,k})+\sum_{k=1}^\infty A_c(t)\alpha_{c,k}\cos(2\pi k\phi_c(t)+\beta_{c,k})\,, 
\]
with $\alpha_{r,k},\alpha_{c,k}\geq 0$ and $\beta_{r,k},\beta_{c,k}\in [0,2\pi)$. We refer to $\alpha_{c,k}\cos(2\pi k\phi_c(t)+\beta_{c,k})$ as the $k$-th harmonic of the cardiac component, and $\phi'_c(t)$ as the instantaneous heart rate (IHR). 
A successful decomposition of the cardiac component should at least reflect the IHR. Unlike PDU, the black crosses in the figure, which mark the R-peak timings from the simultaneously recorded electrocardiogram, align well with the oscillations of the first component extracted by the windowed PDU. This alignment indicates that the cardiac component is successfully extracted, and the mode-mixing issue is mitigated. Additionally, the second component appears sinusoidal and shows two cycles per R-peak, suggesting recovery of the second harmonic. In contrast, while the third component exhibits three cycles per R-peak, its waveform deviates from a clean sinusoid, indicating it does not faithfully capture the third harmonic. Although the theoretical understanding of how noise affects PDU remains incomplete, especially regarding its influence on windowed PDU, we hypothesize that the observed phenomenon likely originates from noise in the PPG signal. 
It is worth noting that a successful decomposition of the cardiac component into its harmonics enables further investigation of hemodynamic properties. For instance, previous studies have shown that phase discrepancies between harmonics can serve as early indicators of hemorrhage \cite{alian2023amplitude}. 
While we do not show the result, when $T$ is large like $T=5$, the decomposed components is not physiologically meaningful like those obtained with PDU.

\begin{figure}[hbt!]
    \centering
\includegraphics[trim=0 0 0 0, clip, width=\columnwidth]{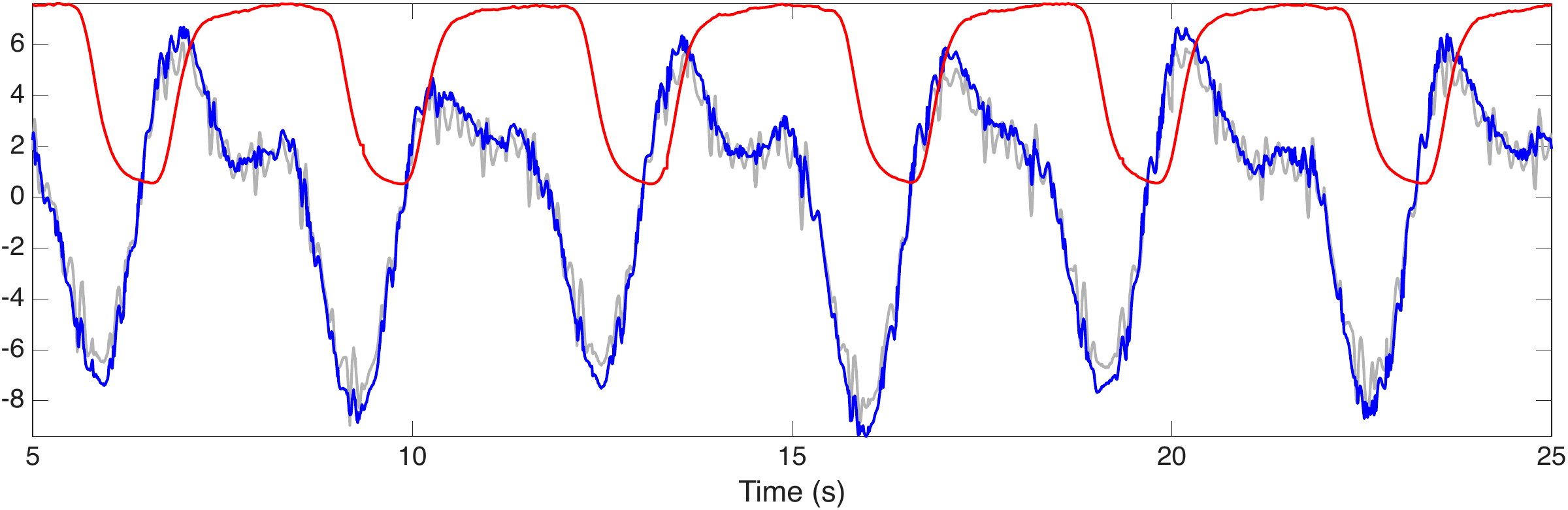}
    \caption{The black curve is the estimated respiratory signal using the windowed PDU and the red curve is the simultaneously recorded end-tidal CO2 signal.}
    \label{fig illustration windowPDU vs PDU PPG-resp}
\end{figure}

Another indication of successful cardiac decomposition is the recovery of the respiratory component. By subtracting the sum of the first $k$ windowed PDU components (i.e., the estimated cardiac component) from the original PPG signal, we obtain an estimate of the respiratory signal. Figure \ref{fig illustration windowPDU vs PDU PPG-resp} displays this result: the gray curve is the estimated respiratory component with $k=3$, the blue curve is the estimated respiratory component with $k=6$ and the red curve shows the simultaneously recorded end-tidal CO2 signal. With $k=6$, the estimated respiratory signal appears smoother than that obtained with $k=3$. While the waveforms differ in morphology, the estimated respiratory signal clearly captures the correct oscillatory pattern.

\bibliographystyle{plain}
\bibliography{ref}

\appendix

\section{Existing theoretical results}\label{section: theoretical results}

{
This section reviews existing theoretical results on the PDU algorithm developed under various mathematical models and from different perspectives. These results can be grouped into four categories, which are summarized in the following four subsections.}

\subsection{Convergence and compression}
The first category concerns the convergence of PDU. The convergence of PDU in   $H^p$, was given in \cite{qian2010intrinsic}. The $L^2$ ($W^{s-1/2}$ in general) convergence of analytic functions without roots on $\partial\mathbb{D}$ in the Dirichlet space ($W^{s}$ in general, where $W^{s}$, $s\in \mathbb{R}$, is the Sobolev space on $\partial \mathbb{D}$.) is given in \cite[Theorem 1]{coifman2017nonlinear}. The $L^p$ convergence, where $p\in (1,\infty)$ is given in \cite[Theorem 1]{coifman2019phase} when the input signal is in $H^p(\partial \mathbb{D})$. Geometric exploration of PDU, including the winding number, is discussed in \cite{coifman2017nonlinear} as well.
Another approach is assuming the function $f$ to be a polynomial of degree $n$. In this framework, the expansion is exact after $n$ steps \cite{Nahon:2000Thesis}, and a further exploration is recently reported in \cite{lukianchikov2019iterative}. 
In \cite{mnatsakanyan2022almost}, under extra mild assumptions on the zeros, the authors prove the $L^p$ bounds for the maximal partial sum operator of the Malmquist-Takenaka series, which generalizes Carlson-Hunt theorem to this setup. 
Numerically, it is commonly observed that PDU converges ``exponentially'' fast, meaning that we usually only need few decomposed components to well recover the original signal. This is the foundation of designing an efficient compression algorithm in practice \cite{tan2018novel}. However, so far only limited results are available under very special assumptions.

\subsection{Decomposition and IF}\label{section: decomposition and IF results}

The second category concerns the analysis of the signal's oscillatory behavior, for example, how many oscillatory components are there in the signal, how fast and how large each component oscillates, and how to decompose the signal into those oscillatory components. For these purposes, we need more specific models. A widely considered model is 
the {\em adaptive harmonic model} (AHM) \cite{DaLuWu2011}. In the original AHM considered in \cite{DaLuWu2011}, the signal is composed of multiple oscillatory components and each component oscillates with time-varying frequency and amplitude, 
\[
f(t)=\sum_{l=1}^LA_l(t)\cos(2\pi\phi_l(t))\,, 
\]
where $t\in \mathbb{R}$, $\phi_l\in C^2$ is strictly monotonically increasing modeling the $l$-th component's phase, $\phi'_l(t)$ models the instantaneous frequency (IF), and $A_l\in C^1$ is positive modeling the amplitude modulation (AM). We call $A_l(t)\cos(2\pi\phi_l(t))$ the $l$-th {\em intrinsic mode type} (IMT) function, which models the $l$-th source in the system under observation. When $L>1$, we assume $\phi'_{l}(t)-\phi'_{l-1}(t)>\Delta$ for all $l=2,\ldots L$ and $\phi'_1(t)>\Delta$ for all $t\in \mathbb{R}$ for some $\Delta>0$ that models the spectral gap. This assumption reminds us of the channel assess method, frequency division multiple access, in the communication system, except that the nonsinusoidal oscillations introduce spectral overlaps. 
The identifiability of the AHM function can be found in \cite{Chen_Cheng_Wu:2014}.
The ultimate objective is to develop a version of the PDU algorithm capable of accurately decomposing $f$ into its IMT component, $f_l$, $l=1,\ldots, L$.  

It has been shown in \cite[Theorem 2.2]{Coifman_Steinerberger_Wu:2016} that a complex-valued periodic function on $[0,1]$ satisfying the AHM conditions; for example, $f(t)=A(t)e^{i2\pi\phi(t)}$ with positive $A(t)$ and a monotonically increasing real function $\phi(t)$ satisfying the regularity and slowly varying conditions of the AHM, can be well approximated by an {\em analytic function} \cite{Gabor:1946,vanderPol:1946,Picinbono:1997} via the Hilbert transform. This result is related to the Nuttall's Theorem \cite{Nuttall:1966} about the quadrature approximation to the Hilbert transform of modulated signals.

We have empirically observed \cite{Coifman_Steinerberger_Wu:2016} that PDU can successfully decompose $f(t)$ satisfying the AHM into its IMT functions even when $\Delta$ is small, provided that certain conditions hold; for example, $A_k(t)\gg \sum_{l=k+1}^LA_l(t)$ for all $t$, $k=1,\ldots L-1$, and each $A_l(t)$ can be well approximated by a polynomial of sufficiently low order. A special polynomial case of this empirical observation is established in \cite[Proposition 3.2]{Coifman_Steinerberger_Wu:2016}. The capability to handle small $\Delta$ is an important advantage of PDU. In contrast, window-based TF analysis methods are typically limited by spectral interference when $\Delta$ is small and the window is not sufficiently large. However, a complete theoretical justification of this property remains open.

\subsection{Noise analysis}

The third category is noise analysis. In practice, noise is inevitable, and we can model noisy signal as a realization of a random process satisfying additive noise model
$Y(t)=f(t)+\Phi(t)$ or multiplicative noise model $Z(t)=\Phi(t)f(t)$,
where $f(t)$ is an analytic function and $\Phi$ is a random process with $0$ mean and finite variance. To our knowledge, the analytic behavior of $Y$ and $Z$ under PDU has not yet been well explored, except some recent results. In \cite[Theorem 3.1]{Coifman_Steinerberger_Wu:2016}, the white noise under Poisson extension from $\partial \mathbb{D}$ to $\mathbb{D}$ is reported, suggesting the stability of PDU under additive white noise. Some numerical exploration reported in \cite{Coifman_Steinerberger_Wu:2016} empirically suggests the robustness of PDU but its robustness behavior is different from other time-frequency analysis tools. To further understand the impact of noise, analyzing how roots are perturbed is necessary, but to our knowledge it is widely open.
In \cite{steinerbergerzeroes}, the robustness is explored from the perspective of root behavior of special random polynomials under the Blaschke decomposition. Let $p_n$ be a random polynomial with $n$ roots that are independently and identically distributed following a probability measure that can be written as $\mu = \phi(\sqrt{x^2+y^2})dx dy$ for some $\phi \in C^{\infty}_c((1,\infty))$. The Blaschke decomposition of $p_n-p_n(0)= B G$ satisfies that $G$ is a random polynomial whose roots are also distributed according to $\mu$ as $n \rightarrow \infty$. Moreover, for typical $n$ degree polynomials where $n$ is large, one cannot expect more than $o(n)$ roots inside the unit disk. 
This theorem suggests that when $n$ is sufficiently large, the Blaschke unwinding series reduces to a simple power series expansion. A similar phenomenon was already observed to occur for functions whose power series expansion has exponentially decaying coefficients in \cite[Proposition 3.2]{Coifman_Steinerberger_Wu:2016}. It seems likely that polynomials with roots outside the unit disk exhibit exponentially decaying coefficients at least in the generic case, and simple power series expansion then naturally leads to exponentially convergence in the unit disk.

\subsection{Relationship with multiscale signal process and deep neural network}

The fourth category is about its relationship with multiscale signal processing and deep neural network.
A deeper look at PDU reveals its strong connection to multiscale signal processing, particularly the early analytic efforts of Grossmann, Morlet, and Meyer using the phase of Hardy functions. Their analysis of holomorphic functions via Cauchy wavelets was among the original inspirations for wavelet theory, highlighting how phase information could uncover singularities and dynamic structures, including phenomena frequently seen in biomedical signals as instantaneous frequency jumps. 
Kronland-Martinet, Morlet, and Grossmann \cite{kronland1987analysis} explored this direction further by leveraging phase and amplitude variability of holomorphic functions to analyze musical transitions. However, they ran into practical limitations in computation and generalization. Later on, this idea was mostly bypassed by the development of orthogonal wavelet transforms. 
PDU revitalizes these ideas by efficiently harnessing both phase and amplitude of holomorphic functions, showing their power in nonlinear biomedical time series analysis. Building on this, we explore how nonlinear complex analysis, particularly function composition and factorization using the root dynamics of functions, can inspire new forms of holomorphic wavelets or waveforms with links to multi-structured dynamical systems.
In particular, we construct generalized, scaled holomorphic orthogonal bases associated with dynamical systems defined by Blaschke factor compositions, capturing the analytic and geometric role of phase. Such idea has led to a detailed construction of a Malmquist-Takenaka basis for multiscale analysis on $\hardyR{2}$ in \cite{coifman2022multiscale}. See \cite{coifman2019phase} as well. While various constructions exist, a key question is identifying the ``best'' basis. Intuitively, an ideal one should be adapted to the signal via PDU. By iteratively removing oscillations through Blaschke division, we aim to build a natural orthogonal expansion aligned with a Malmquist-Takenaka basis, though this adaptive direction remains largely unexplored beyond the greedy methods proposed in adaptive Fourier analysis \cite{qian2011algorithm}.

In the same paper \cite{coifman2022multiscale}, the authors made a striking theoretical observation that the phase of a Blaschke factor behaves like a {\em one-layer neural network} with the activation function $\arctan$. Consequently, compositions of Blaschke factors resemble a deep neural network, where the depth corresponds to the number of compositions. To take a closer look,  take $\{a_j=\alpha_j+i\beta_j\}_{j=1}^n\subset\mathbb{C}_+$ and the associated Blaschke product on the real line is
\begin{align*}
\blp(x) = \prod_{j=1}^n \frac{x-a_j}{x-\cnj{a}_j}=\exp(i\theta(x))\,,
\mbox{ where }
\theta(x)=\sum_{j=1}^n\sigma\Big(\frac{x-\alpha_j}{\beta_j}\Big)
\end{align*} 
and $\sigma(x)=\arctan(x) + \pi/2$ is a sigmoid \cite{coifman2022multiscale}.
Each term in $\theta(x)$ mimics a neuron with its own weight and bias, and the sum forms a single layer. Deeper networks arise naturally from further compositions of Blaschke products.
This neural-network-like structure parallels the iterative schemes studied by Daubechies et al. \cite{daubechies2022nonlinear}, where piecewise affine maps play a similar compositional role to Blaschke products.

\subsection{Relationship to similar ideas}

A natural approach to to capture the dynamics when the time series oscillates with time-varying amplitude or frequency is analyzing the time series by reading its local spectral behavior via the ``divide-and-conquer'' approach; that is, analyzing the time series from the time and frequency perspectives simultaneously. This idea leads to the TF analysis. Examples range from the widely applied short time Fourier transform (STFT), continuous wavelet transform (CWT), and Wigner-Ville distribution (WVD), ad hoc empirical mode decomposition, to recently developed synchrosqueezing transform, among many others. We refer readers to a recent review article \cite{wu2020current}.
Due to the nonlinear dynamics of the roots involved in the PDU algorithm, it is categorized as a nonlinear TF analysis algorithm. By analogy, PDU relates to Fourier analysis as a musical score (and/or orchestration) relates to a Fourier expansion, where the score is a transcription of notes of different duration at a given pitch at different instances (zeroes of a holomorphic signal are orchestrated in PDU), while Fourier spectrum consists of unlocalized notes of preassigned fixed period. 
Altogether, due to its intimate relationship with the Fourier series, and its ability to capture the notion of IF, PDU is an extension of the Fourier transform useful address nonlinear and nonstationary time series.

We mention that an idea similar to PDU was considered in \cite{qian2010intrinsic} based on the Malmquist-Takenaka system \cite{takenaka1925orthogonal,eisner2014discrete}, where the authors called the algorithm {\em adaptive Fourier decomposition} \cite{qian2011algorithm}. The adaptive Fourier decomposition is a greedy {iterative} algorithm in nature. Instead of generating zeros by subtracting $G_l(0)$ in the $l$-th step, in the adaptive Fourier decomposition the user subtracts $G_l(\alpha_l)$ in the $l$-th step, where $\alpha_l\in \mathbb{D}$ so that $G_l(\alpha_l)$ has the most energy. 
Empirically, its performance in the sense of convergence rate is similar to the original PDU algorithm, and since $G_l(\alpha_l)$ is a constant, it is also limited in capturing AM. 
It is possible to consider other general orthogonal functions \cite{pap2006voice,feichtinger2013hyperbolic} to achieve the decomposition. A more general formulation of PDU based on invariant subspaces is presented in \cite{coifman2019phase}. We refer interested readers to the literature for further details.

The relationship between PDU and adaptive Fourier decomposition naturally suggests a connection to the Mobius transform. One can move the point $\alpha_l$ carrying the most energy to $0$ considered in adaptive Fourier decomposition by considering $G_l\circ M_{\alpha_l}$, where $M_{\alpha_l}(z)=\frac{z+\alpha_l}{1+\bar{\alpha_l}z}$, and then applying PDU to $G_l\circ M_{\alpha_l}$. This procedure is equivalent to performing a Blaschke decomposition on $G_l\circ M_{\alpha_l}-G_l\circ M_{\alpha_l}(0)=G_l\circ M_{\alpha_l}-G_l(\alpha_l)$. We defer investigation of this relationship via Mobius transform to future work.

\end{document}